\documentclass[12pt]{article}
\usepackage{amssymb,amsmath,amsthm,epsfig,bm}
\numberwithin{equation}{section}
\numberwithin{figure}{section}
\usepackage{epic,graphicx,color}
\usepackage{ifpdf}
\usepackage{mathrsfs}
\usepackage{chemformula}

\usepackage[noend]{algpseudocode}
\usepackage{algorithmicx,algorithm}

\usepackage{multirow}
\usepackage{makecell}
\usepackage{booktabs}
\usepackage{dsfont}
\usepackage{tabularx}
\usepackage{diagbox}
\usepackage{caption,subcaption}

\usepackage{threeparttable}
\usepackage[numbers,sort&compress]{natbib}

\theoremstyle{plain}

\theoremstyle{definition}

\theoremstyle{remark}
\newtheorem{rem}{Remark}[section]

\usepackage{indentfirst}

\begin{document}
	
	\title{ \large\bf MsFEM-Inspired CNNs with Transfer Learning \\ for Multiscale Model Reduction}
	
	\author{Xuehan Zhang\thanks{School of Mathematical Sciences,  Tongji University, Shanghai 200092, China. ({\tt  xhzhang@tongji.edu.cn}).}
		\and
		Lijian Jiang\thanks{School of Mathematical Sciences, Key
			Laboratory of Intelligent Computing and Applications (Ministry of Education),  Tongji University, Shanghai 200092, China. ({\tt  ljjiang@tongji.edu.cn}).}
		\and
		Eric T. Chung\thanks{Department of Mathematics, The  Chinese University of Hong Kong, Hong Kong 999077, China. ({\tt  tschung@math.cuhk.edu.hk}).}
	}
\date{}

\maketitle
\begin{center}{\bf Abstract}
\end{center}\smallskip

Deep learning-based surrogate models have been extensively developed for efficiently approximating multiscale systems with random input fields. However, most existing approaches require retraining neural networks from scratch when source terms, boundary conditions, or differential operators change, resulting in significant computational costs and limited adaptability. To address this challenge, we integrate our previous CNN-based reduced-order model (ROM) framework with the multiscale finite element method (MsFEM) and propose an MsFEM-inspired transfer learning strategy, termed MITL. The CNN-based ROM consists of two components: Basis CNNs, which learn reduced basis functions, and Coef CNNs, which predict the corresponding linear combination coefficients. To enhance the transferability of learned multiscale representations, global MsFEM basis problems are employed as source tasks during pretraining. For new target problems, MITL requires training only lightweight adaptation networks to construct task-specific reduced bases and coefficients, thereby substantially reducing the computational burden. Numerical experiments demonstrate that MITL achieves accurate and efficient predictions across a range of target tasks, with particularly significant advantages in data-scarce scenarios.

\smallskip
{\bf keywords: transfer learning; convolutional neural network; multiscale finite element method;} 

\section{Introduction}
\label{sec1:introduction}
Multiscale problems are widely studied across many science and engineering fields, such as materials science, environmental science, biomechanics, and so on. These problems are typically modeled by partial differential equations (PDEs) with multiscale random input fields. The randomness usually arises from the heterogeneity of physical properties. For instance, the hydraulic conductivity field in groundwater flow is often highly heterogeneous and exhibits random multiscale variations due to the spatial distribution of pores, fractures, and soil layers with different permeabilities. Such heterogeneity typically leads to multiscale coefficients in the governing  flow equations \cite{sec1:Darcy_flow_porous}. To capture the fine-scale features of such systems, numerical simulation methods must solve large-scale algebraic equations, which is  often computationally prohibitive. Therefore, it is necessary to design efficient multiscale simulation methods, especially for real-time applications (e.g., inverse problems and stochastic control \cite{sec1:inverse problems,sec1:optimal control}) and many-query contexts (e.g., design or optimization \cite{sec1:multiscale design}).

During last decades, a few reduced-order modeling methods (ROMs) have been proposed for multiscale problems. ROMs can be classified as intrusive or non-intrusive, depending on whether they require the underlying equations. They can all be summarized into two stages: offline stage and online stage. In the offline stage, a set of reduced basis is constructed using using either data-driven or physics-driven methods from snapshots of high-fidelity responses. Data-driven approaches extract global and general features from snapshots through techniques such as proper orthogonal decomposition (POD) \cite{sec1:POD-based ROM} and greedy algorithms \cite{sec1:greedy-based ROM}), while physics-driven methods (e.g., multiscale finite element methods \cite{sec1:MsFEM1,sec1:MsFEM2,sec1:MsFEM3} and its variants \cite{sec1:GMsFEM1,sec1:CEM-GMsFEM1,sec1:Splitting}) capture multiscale features directly from local multiscale operators for each input sample. Then, in the online stage, the linear combination of the reduced basis can be efficiently determined by either Galerkin projection (intrusive)\cite{sec1:intrusive_ROM1,sec1:intrusive_ROM2,sec3_CNNbasedROM} or regression methods (non-intrusive) \cite{sec1:DeepGlobalModelReduction,sec1:ROMs1,sec1:ROMs2,sec1:ROMs3,sec1:ROMs4,sec1:ROMs5}. 

By taking the advantages of intrusive reduced-basis methods and multiscale model reduction methods, we proposed a deep model reduction methods, called CNN-based ROM in \cite{sec3_CNNbasedROM}, for multiscale problems. It consists of two CNNs: Basis CNNs and Coef CNNs. They corresponds to the two phases  of ROMs. The first one is used to learn reduced basis from data constrained by equations, and the second one is designed to learn the coefficients of linear conbinations. Various numerical experiments demonstrate that CNN-based ROM can serve as an efficient surrogate for multiscale problems with random input fields. However, the CNN-based ROM suffers from three limitations that hinder its applicability to real-world problems. First, it is an intrusive ROM, as the training of the Basis CNNs relies on the underlying governing equations. Second, a large amount of high-fidelity data is required to avoid overfitting. Third, any changes in the source term, boundary conditions, or differential operators necessitate retraining the model, which is a common drawback of AI-for-PDE methods  \cite{sec1:TL1,sec1:TL2,sec1:TL3}.

In this paper, we consider the situation where  the underlying equations are only partially known. Moreover, the size of dataset is limited, and the data can be obtained through numerical simulations or physical experiments. This work aims to develop an efficient transfer learning strategy grounded in CNN-based ROM. The strategy is designed to transfer multiscale features from source tasks to target tasks. Thus, we can achieve rapid adaptation by training lightweight neural networks. Inspired by the fact that the multiscale basis of MsFEM is transferable to problems with different source terms and boundary conditions, we propose the transfer learning strategy, called MsFEM-inspired CNNs with transfer learning (MITL). The global MsFEM problems are chosen as source tasks to obtain transferable multiscale features, which is trained by CNN-based ROM. Corresponding to two CNNs in CNN-based ROM, the target model also consists of two CNNs: TL-Basis CNNs and TL-Coef CNNs. TL-Basis CNNs keep the parameters of Basis CNNs fixed and add a residual structure with a two-layer convolutional neural operator (CNO) to obtain the new reduced basis for target tasks. In TL-CoefCNNs, only the top-layer parameters of Coef CNNs are trainable. Additionally, dimension-modification layers are designed for adjusting the output dimension.  Furthermore,  time encoding layers are proposed for solving unsteady problems. 

This article is structured as follows. In Section \ref{sec2:problem setup}, we give the definitions and examples of multiscale problems, which is followed by a brief introduction to CNN-based ROM and MsFEM in Section \ref{sec3:preliminary}. Section \ref{sec4:methodology} focuses on the proposed method, MITL. In this section, we also explore the connections between the proposed method with numerical homogenization. In Section \ref{sec5:numerical experiment}, a few numerical results are presented to illustrate the efficacy of the proposed method. We also demonstrate that MITL can provide  a significant  surrogate model for efficient  inversion. Finally, some conclusions are given.

\section{Problem setup}
\label{sec2:problem setup}
In this paper, we consider the following multiscale boundary value problems defined in a bounded domain $\Omega\subset\mathbb{R}^2$ with boundary $\partial{\Omega}$:
\begin{equation}
	\left\{
	\begin{aligned}
		&\mathcal{L}\Big(u(x);\kappa(x,\xi)\Big)=f(u,x),\quad x \in \Omega,\\
		&\mathcal{D}u(x)=b(x),\quad x \in \partial{\Omega},\\
	\end{aligned}
	\right.
	\label{sec2_eq:problem setup}
\end{equation}
where $\mathcal{L}$ denotes a nonlinear differential operator. The randomness of above system is described by the parametric random field  $\kappa(x,\xi)$, which depends explicitly on the spatial variable $x\in \Omega$ and the $d-$dimensional random vector $\xi \in (\Xi,\mathcal{F},\mathbb{P})$. This random field characterizes the multiscale features of the medium. In numerical experiments, samples of $\kappa(x,\xi)$ are obtained either through direct sampling or from real observational data. The term $f(u,x)$ is the nonlinear source term with sufficient regularity. The boundary condition operator $\mathcal{D}$ can represent Dirichlet, Neumann, or mixed boundary conditions, and $b(x)$ denotes the given boundary condition (e.g., the Dirichlet trace or Neumann flux).

Although many existing approaches can provide efficient surrogate models for above problems, they require constructing separate models and retraining from scratch when the source terms, boundary conditions, or differential operators vary. Therefore, our goal is to develop effective transfer learning strategies that enable rapid transfer across similar multiscale problems. In detail, to learn transferable multiscale features, we first select a specific instance of the above problem as the source task, namely,
\begin{equation}
	\left\{
	\begin{aligned}
		&\mathcal{L}_s\Big(u_s(x);\kappa(x,\xi)\Big)=f_s(u,x),\quad x \in \Omega,\\
		&\mathcal{D}_su_s(x)=b_s(x),\quad x \in \partial{\Omega}.\\
	\end{aligned}
	\right.
	\label{sec2_eq:source tasks}
\end{equation}
The formulation of the source task (\ref{sec2_eq:source tasks}) is assumed to be known and captures all the multiscale features. Let $\Omega$ be partitioned by a uniform grid with mesh sizes $h$ (denoting the spatial grid by $\mathcal{H}$). And $K\in \mathbb{R}^{N_h \times N_h}$ is the sample of $\kappa(x,\xi)$, which takes values on $\mathcal{H}$. Then, we can obtain finite-dimensional approximations $u_s^h \in V_s^h \subset \mathbb{R}^{N_h \times N_h}$ of $u_s(x)$ through a finite element method (FEM). The stiffness matrix and load vectors of FEM are denoted by $A_h(u_s^h;K)$ and $F_h(u_s^h;K)$. Therefore, we can construct a surrogate for the source task 
\begin{equation*}
	u_s^h = g_s(K),
	\label{sec2_eq:source model}
\end{equation*}
using source dataset $\boldsymbol{D} = \Big\{K_i,u_s^h(K_i),A_h(u_s^h;K_i),F_h(u_s^h;K_i)\Big\}_{i=1}^{N_s}.$ Here  the  source  neural network   $g_s$ : $K\longrightarrow V_s^h$.

Similarly, we consider a target task that differs from the source task in one or more components (e.g., the differential operator $\mathcal{L}_t$, the source term $f_t$, or the boundary data $b_t$), while sharing the same parametric random field $\kappa(x,\xi)$ and the same spatial domain $\Omega$:
\begin{equation}
	\left\{
	\begin{aligned}
		&\mathcal{L}_t\Big(u_t(x);\kappa(x,\xi)\Big)=f_t(u,x), \quad x \in \Omega,\\
		&\mathcal{D}_t u_t(x)=b_t(x), \quad x \in \partial\Omega.
	\end{aligned}
	\right.
	\label{sec2_eq:target tasks}
\end{equation}
The explicit form of target task (\ref{sec2_eq:target tasks}) is unkown and only limited data are available. For any input sample $K$, $u_t(x)$ is observed on the grid $\mathcal{H}$, leading to the matrix $u_t^h(K)$. And then we can obtain the following target model by using target dataset $\boldsymbol{D}_t = \Big\{K_i,u_t^h(K_i)\Big\}_{i=1}^{N_t},N_t\ll N_s$,
\begin{equation*}
	u_t^h = g_t(K),
	\label{sec2_eq:target model}
\end{equation*}
where $g_t$ retains the multiscale features acquired by $g_s$ and only lightweight neural networks in target model are trained to adapt to target tasks.

\section{Preliminaries}
\label{sec3:preliminary}
In this section, we introduce the necessary preliminaries for the proposed method. We first provide a brief review of our previous work on CNN-based ROM  \cite{sec3_CNNbasedROM}. Subsequently, Subsection \ref{ssec:MsFEM} revisits MsFEM from a transfer learning perspective, which serves as the primary inspiration for our approach. We also discuss the limitations of these two methods, which motivate the development of the proposed method.

\subsection{CNN-based ROM}
\label{ssec:CNNbasedROM}
\begin{figure}[htp]
	\centering
	\includegraphics[scale=0.3]{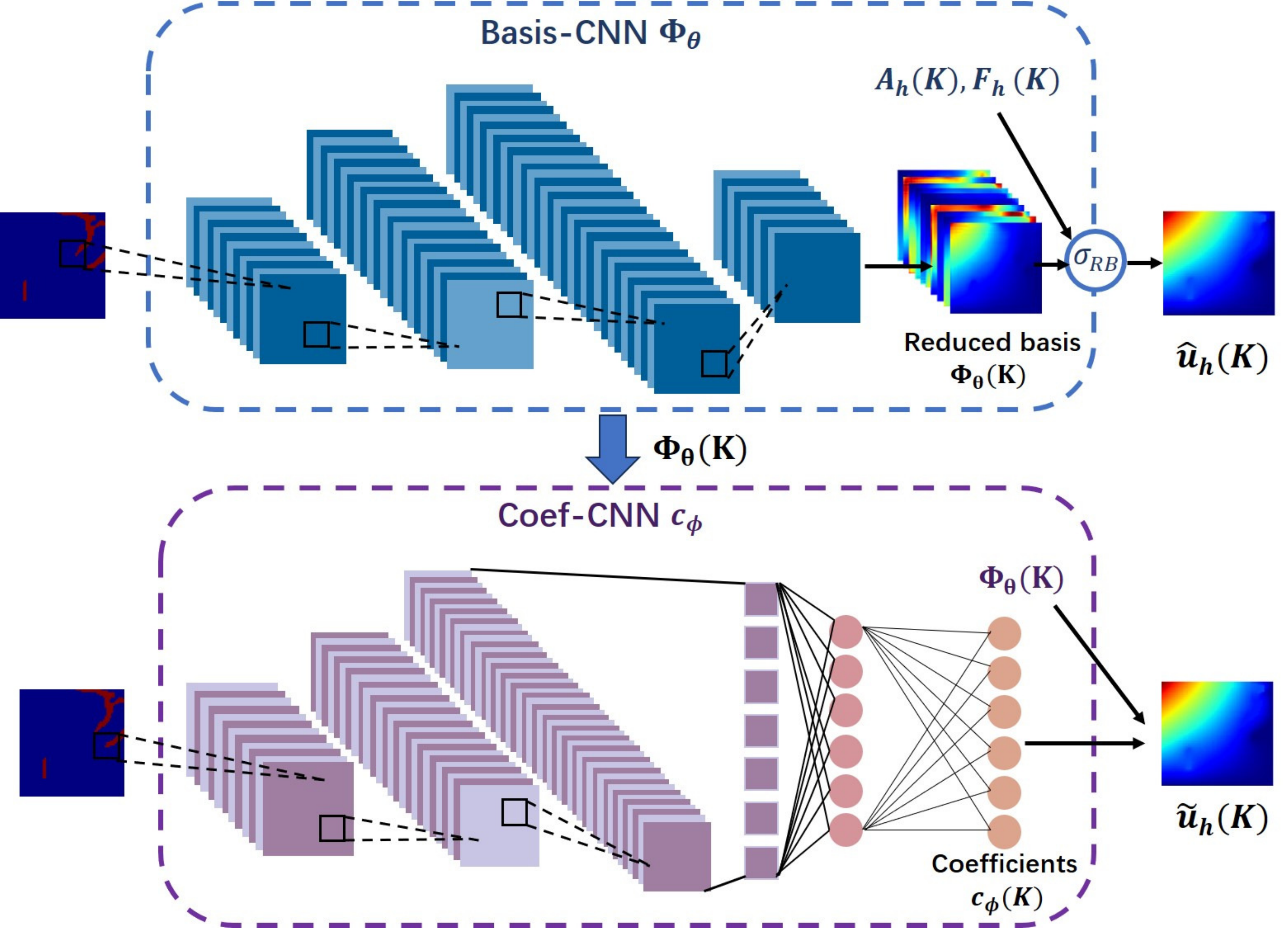}
	\caption{The schema of CNN-based ROM.}	
	\label{sec3_fig:CNNbasedROM}
\end{figure}

 As illustrated in Figure \ref{sec3_fig:CNNbasedROM}, CNN-based ROM consists of two separate CNNs: Basis CNNs and Coefficient CNNs (Coef CNNs), which correspond to two main parts of ROMs. The first
one learns input-specific basis functions from the snapshots of fine-scale solutions. An activation function, inspired by Galerkin projection, is utilized at the output layer to reconstruct fine-scale solutions from the basis functions. Given the dataset
 \begin{equation*}
 	\Big\{K_i,u_h(K_i),A_h(u_h;K_i),F_h(u_h;K_i)\Big\}_{i=1}^{M},
 \end{equation*}
 a low-dimensional approximation of fine-scale solution $u_h$ can be written as,
\begin{equation*}
	\hat{u}_h(K) = \Big(\Phi_{\theta}(K)^TA_h(u_h;K)\Phi_{\theta}(K)\Big)^{-1}\Big(\Phi_{\theta}(K)^TF_h(u_h;K)\Big),
\end{equation*}
where $\Phi_{\theta}(K)\in \mathbb{R}^{N_h^2\times N}$ is the learned reduced basis, each column of which represents a reduced base. Then the parameters $\theta$ of Basis CNN  can be optimized by,
\begin{equation*}
	\arg\min_{\theta}\frac{1}{M}\sum_{i=1}^M \Vert u_h(K_i)-\hat{u}_h(K_i)\Vert_2^2+\text{cond}_F\Big(\Phi_{\theta}(K)^TA_h(u_h;K)\Phi_{\theta}(K)\Big),
\end{equation*}
where $\text{cond}_F$ denotes the matrix condition number under the Frobenius norm. It should be noted that the tests of Basis CNNs still require the fine-scale stiffness matrix and load vector, which limits online computational efficiency. More importantly, they cannot be directly applied to nonlinear problems. The second CNN, called Coef CNN, is then designed to determine the coefficients for learning the linear combination of reduced basis, namely,
\begin{equation*}
	\tilde{u}_h(K) = \Phi_{\theta}(K)c_{\phi}(K),
\end{equation*}
where Coef CNN $c_{\phi}(K)$ defines a mapping from input sample $K$ to cofficients. The parameters $\phi$ of Coef CNN can be optimized by the Mean Squared Error (MSE) loss function,
\begin{equation*}
	\arg\min_{\phi}\frac{1}{M}\Vert u_h(K_i)-\tilde{u}_h(K_i)\Vert_2^2
\end{equation*}
Various numerical experiments demonstrate that the proposed method achieves good performance in both approximation accuracy and generalization. Therefore, it is a suitable choice as the source model for transfer learning tasks. However, there are also some limitations in real-world applications.
\begin{itemize}	
	\item The training of Basis CNN relies on data, and the solution data inherently encodes information such as the source terms and boundary conditions. As a result, the learned reduced basis cannot achieve effective transfer across different source terms, boundary conditions, unlike those in the multiscale finite element method.
	
	\item The training of Basis CNN relies on access to the underlying governing equations, which are used to constrain the learning of reduced basis functions. For practical multiscale problems, the exact form of the equation is often unavailable, thereby limiting the applicability of the method.
	
	\item The training of CNN-based ROM typically requires a large number of high-fidelity snapshot datasets. However, high-accuracy numerical simulations for multiscale problems demand very fine mesh discretizations, making the offline data generation process computationally expensive and limiting its applicability to large-scale engineering problems.
\end{itemize}

\subsection{Insights into MsFEM from a Transfer Learning Perspective}
\label{ssec:MsFEM}
The main idea of transfer learning is to leverage knowledge learned from source tasks to improve learning efficiency on a target task. This is typically achieved by extracting and reusing transferable representations that capture shared features across tasks, enabling faster adaptation and reducing the need for training from scratch. For example, in the field of image classification, a deep neural network pre-trained on a source domain task often learns general features unrelated to the specific task (such as low-level patterns like edges and textures) in its shallow layers, while its deeper layers capture task-relevant semantic information. By freezing the shallow-layer parameters of the pre-trained model and fine-tuning the deep-layer parameters, the general knowledge acquired from the source domain can be transferred to the target task.

In this subsection, we review MsFEM from the perspective of transfer learning, summarizing its methodology as the following transfer learning framework.
\begin{itemize}
	\item \textbf{Source task:} First, a source model is selected, which defines a finite-dimensional projection of the full-order solution,
	\begin{equation*}
		g_s: V^{N_h} \rightarrow V^N, \quad \hat{u}_{h} = \Phi_{\text{ms}} c_{\text{ms}},
		\label{sec3_eq:msfem_sourcemodel}
	\end{equation*}
	where $\Phi_{\text{ms}} = [\phi_{\text{ms}}^1, \cdots, \phi_{\text{ms}}^N]$ denotes the multiscale basis functions and $c_{\text{ms}}$ represents the reduced-order coefficients. Subsequently, suitable local boundary conditions $\mathcal{D}_{\text{ms}}$\cite{sec1:MsFEM1} are imposed. The basis functions $\Phi_{\text{ms}}$, which capture the multiscale features, are then obtained by solving the following local problems,
	\begin{equation}
		\left\{
			\begin{aligned}
				&\mathcal{L}_{\text{ms}}\big(\phi_{\text{ms}}^i(x); \kappa(x,\xi)\big) = 0,\quad x \in \Omega_k, \\
				&\mathcal{D}_{\text{ms}}^{i} \phi_{\text{ms}}^i(x) = b(x),\quad x \in \partial \Omega_k, i=1,\cdots,N_{ms},
			\end{aligned}
			\right.
			\label{sec3_eq:local_msfem}
	\end{equation}
	where $\Omega_k$ is a local subdomain of $\Omega$ such that $\Omega = \bigcup \Omega_k$.
	
	\item \textbf{Target task:} For a new source term $f(x)$ or boundary conditions $\mathcal{D}$, 
	\begin{equation}
		\left\{
		\begin{aligned}
			&\mathcal{L}_{\text{ms}}\big(u(x); \kappa(x,\xi)\big) = f(x),\quad x \in \Omega, \\
			&\mathcal{D} u(x) = b(x),\quad x \in \partial \Omega,
		\end{aligned}
		\right.
		\label{sec3_eq:MsFEM_target}
	\end{equation}
	the multiscale basis $\Phi_{\text{ms}}$ obtained from the cell problems (\ref{sec3_eq:local_msfem}) are treated as the transferable component and remain fixed. The reduced-order coefficients $c_{\text{ms}}^t$ are then determined via Galerkin projection,
	\begin{equation*}
		c_{\text{ms}}^t = (\Phi_{\text{ms}}^TA_h\Phi_{\text{ms}})^{-1}\Phi_{\text{ms}}^TF_h,
	\end{equation*}
	where $A_h$ and $F_h$ are stiffness matrix and load vector of the target task (\ref{sec3_eq:MsFEM_target}). 
	\end{itemize}
	
	In contrast to purely data-driven transfer learning, feature transfer in MsFEM is grounded in physical principles rather than relying on the empirical extraction of data distributions. This fundamental difference constitutes the two sides of the same coin in this method. On the one hand, its strengths lie in generalization and interpretability. Since the transferred knowledge is derived from underlying physical mechanisms, the basis functions $\Phi_{\text{ms}}$ possess strong theoretical guarantees and stable extrapolation capabilities when applied to different source terms or boundary conditions under the same medium coefficient $\kappa(x,\xi)$.
	
	On the other hand, its transferability exhibits two main limitations:
	\begin{itemize}
		\item The transfer of multiscale basis functions $\Phi_{\text{ms}}$ heavily depends on input sample itself. In practice, they can only be applied to problems with the same random input sample or to samples with slight variations, making it difficult to share features across different inputs as in data-driven models.
		\item When the operator $\mathcal{L}_{\text{ms}}$ changes, especially when it becomes nonlinear, the transferability of the multiscale basis functions lacks theoretical justification and may even break down. 
	\end{itemize} 

	Inspired by the physics-driven transfer paradigm of MsFEM, this work further explores its extension and implementation within a data-driven framework. On the one hand, the core ideas of MsFEM are adopted as a theoretical foundation to guide the design of transfer learning strategies. On the other hand, the nonlinear approximation capability of neural networks and large-scale data pretraining mechanisms are introduced to overcome the limitations of MsFEM basis functions in cross-sample and cross-operator transfer tasks.
	
\section{MsFEM-Inspired CNNs with Transfer Learning}
\label{sec4:methodology}
\begin{figure}[htp]
	\centering
	\includegraphics[scale=0.26]{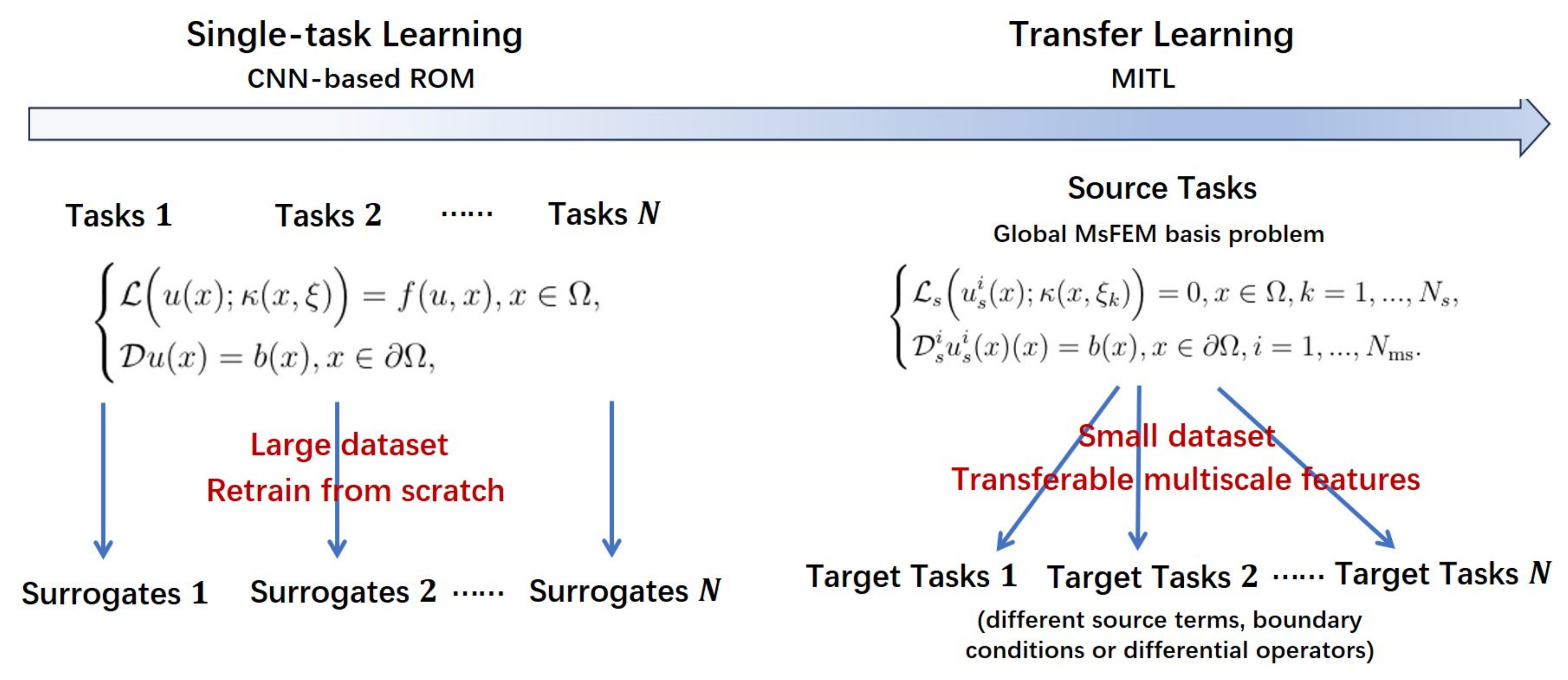}
	\caption{The schema of MITL.}	
	\label{sec4_fig:MITL_outline}
\end{figure}
In Section \ref{ssec:CNNbasedROM}, we analyze the limitations of CNN-based ROM. Inspired by MsFEM, we know that the basis learned from cell problems (\ref{sec3_eq:local_msfem}) does not depend on source terms and can be reused for solving new equations with different source terms and boundary conditions. As illustrated in Section \ref{ssec:MsFEM}, such property resembles transfer learning techniques. 

In terms of the ideas of MsFEM and transfer learning techniques, we propose a novel multiscale model reduction approach, called MsFEM-inspired CNNs with transfer learning (MITL). The proposed method consists of two stages: pretraining on a source task and adaptation to a target task (see Figure \ref{sec4_fig:MITL_outline}). It aims to extract transferable multiscale feature representations from the source tasks and achieve efficient adaptation using only a small amount of target task data.

\subsection{Source task}
Multiscale problems (\ref{sec2_eq:problem setup}) involve a wide variety of source terms and boundary conditions. It is crucial to select a representative subset of problems (\ref{sec2_eq:problem setup}) as the source tasks for constructing MITL. As noted in Section \ref{ssec:MsFEM}, the multiscale basis constructed by  MsFEM have been demonstrated, both theoretically and numerically, to be transferable to problems with different source terms and boundary conditions. Therefore, it is natural to choose the cell problem (\ref{ssec:MsFEM}) as source task of the proposed method.

In MsFEM, the multiscale basis functions are solved locally on subdomains $\Omega_k$. This localization strategy is a trade-off between accuracy and computational efficiency. In other words, solving the cell problems over the whole domain $\Omega$ would lead to computational cost similar to that of the original problems, thereby failing to achieve the purpose of model reduction. However, the multiscale features inherently embedded in the heterogeneous input $\kappa(x,\xi)$ have a global influence on the output response $u_h$ \cite{sec4_nonlocal}. Local truncation inevitably results in the loss of interaction information between different regions. This loss of information leads to the use of a relatively large number of basis functions in MsFEM to achieve the desired approximation accuracy.

To capture more complete multiscale features, the restriction to local formulations is removed, and the source task is defined directly on the global domain. Although global computation are computationally more expensive, this cost remains acceptable as the source task is performed in the offline stage. More importantly, the global formulation enables the multiscale basis to capture the global coupling across scales, thereby allowing the source model to learn more intrinsic and transferable representations. Specifically, the source tasks are posed as the following $N_{\text{ms}}$ global boundary value problems,
\begin{equation}
	\left\{
	\begin{aligned}
		&\mathcal{L}_s\Big(u_s^i(x);\kappa(x,\xi_k)\Big)=0,\quad x \in \Omega,\quad k = 1,\cdots,N_s,\\
		&\mathcal{D}_s^{i}u_s^i(x)(x) = b(x),\quad x \in \partial{\Omega},\quad i = 1,\cdots,N_{\text{ms}}.\\
	\end{aligned}
	\right.
	\label{sec4_eq:source problem1}
\end{equation}

In this work, we take the following linear problems defined on the square domain $\Omega = [x_1,x_2]\times [y_1,y_2]$ as an example,
\begin{equation*}
	\left\{
	\begin{aligned}
		&-\nabla\cdot\bigl(\kappa(x,\xi_k)\nabla u_s^i(x)\bigr)=0,\quad x \in \Omega,\; k = 1,\dots,N_s,\\[4pt]
		&u_s^i(x) \text{ is linear on } \partial\Omega \quad \text{and} \quad u_s^i(x^j)=\delta_{ij},
	\end{aligned}
	\right.
\end{equation*}
where $x^{1} = [x_1,y_1]^T, x^{2} = [x_1,y_2]^T, x^{3} = [x_2,y_1]^T, x^{4} = [x_2,y_2]^T$ and $\delta_{ij}$ is defined as follows,
\begin{equation*}
	\delta_{ij} = 
	\left\{
	\begin{aligned}
		&1,\quad j = i,\\
		&0,\quad j\neq i.
	\end{aligned}
	\right.
\end{equation*}
Since the boundary condition is rotationally symmetric, the four boundary value problems can be transformed into a single equivalent boundary value problem by rotating the random input sample $\kappa(x,\xi_k)$, 
\begin{equation}
	\left\{
	\begin{aligned}
		&-\nabla\cdot\Big(\kappa(x,\xi_k)\nabla u_s(x)\Big)=0,\quad x \in \Omega,\quad k = 1,\cdots,4N_s,\\
		&u_s(x) \text{ is linear on } \partial\Omega \quad \text{and} \quad u_s(x^j)=\delta_{1j}. \\
\end{aligned}
\right.
\label{sec4_eq:source problem2}
\end{equation}
Define the two-dimensional rotation matrix by
\begin{equation}
	\boldsymbol{R}_{\theta} = \begin{pmatrix}
		\cos\theta & -\sin\theta \\
		\sin\theta & \cos\theta
	\end{pmatrix},
	\nonumber
\end{equation}
then the rotation operator can be written as $\boldsymbol{T}_{\theta}f(x) = f(\boldsymbol{R}_{\theta}x)$. This operator represents a clockwise rotation of the function $f$ by an angle $\theta$ in the two-dimensional space. Denote the random input sample functions in the source tasks (\ref{sec4_eq:source problem1}) by $\Big\{\kappa(x,\xi_k)\Big\}_{k=1}^{N_s}$. Then, the other $3N_s$ random input sample functions in the source tasks (\ref{sec4_eq:source problem2}) can be defined as follows,
\begin{equation}
	\kappa(x,\xi_{k+N_s}) = \boldsymbol{T}_{\pi/2}\Big(\kappa(x,\xi_{k})\Big) ,
	\kappa(x,\xi_{k+2N_s}) = \boldsymbol{T}_{\pi}\Big(\kappa(x,\xi_{k})\Big),
	\kappa(x,\xi_{k+3N_s}) = \boldsymbol{T}_{3\pi/2}\Big(\kappa(x,\xi_{k})\Big).
	\nonumber
\end{equation}
The spatial domain $\Omega$ is discretized using a uniform mesh $\mathcal{H}$ with mesh size $h$. The source problem (\ref{sec4_eq:source problem2}) is then solved by the finite element method to obtain the source dataset $\mathcal{D}_s = \Big\{K_i,u_s^i(K_i),A_h(K_i),F_h(K_i)\Big\}_{i=1}^{4N_s}$. Then, the source model can be obtained by training a CNN-based ROM, namely,
\begin{equation}
	\hat{u}_s^h = g_s(K) = \Phi_{\theta}(K)c_{\phi}(K),
	\label{sec4_eq:source model}
\end{equation}
where $\Phi_{\theta}(K)$ and $c_{\phi}(K)$ contain the transferable multiscale features.

\subsection{Target task}
In this section, we use a feature-based transfer learning strategy to efficiently transfer the multiscale features learned from the source model to the target task. This leads to a sample-efficient adaptation to the target task.
The effectiveness of such a strategy, however, relies on the similarity between the source and target tasks. To this end, we consider the following four classes of equations, i.e., (\ref{sec2_eq:2ord linear})–(\ref{sec2_eq:p-laplacian}). 
\begin{itemize}
	\item \textbf{Second-order linear partial differential equation}. 
	\begin{equation}
			\left\{
			\begin{aligned}
					&-\nabla \cdot \Big( \kappa(x,\xi)\nabla u(x)\Big)+c u(x) =f(x),\ x \in \Omega,\\
					&\mathcal{D}u(x) =b(x),\ x \in \partial{\Omega}.
				\end{aligned}
			\right.
			\label{sec2_eq:2ord linear}
		\end{equation}
	\item \textbf{Second-order linear partial differential equation with a nonlinear source term}.
	\begin{equation}
			\left\{
			\begin{aligned}
					&-\nabla \cdot \Big( \kappa(x,\xi)\nabla u(x)\Big) = f(x,u(x)),\ x \in \Omega,\\
					&\mathcal{D}u(x) =b(x),\ x \in \partial{\Omega}.
				\end{aligned}
			\right.
			\label{sec2_eq:2ord linear_nonlinearrighthand}
		\end{equation}
	\item \textbf{P-Laplacian equation}.
	\begin{equation}
			\left\{
			\begin{aligned}
					&-\nabla \cdot \Big( \kappa(x,\xi)| u(x)|^{p-2}\nabla u(x)\Big) =f(x),\ x \in \Omega,\\
					&\mathcal{D}u(x) =b(x), \ x \in \partial{\Omega}.
				\end{aligned}
			\right.
			\label{sec2_eq:p-laplacian}
		\end{equation}
\end{itemize}

\begin{figure}[htp]
	\centering
	\includegraphics[scale=0.22]{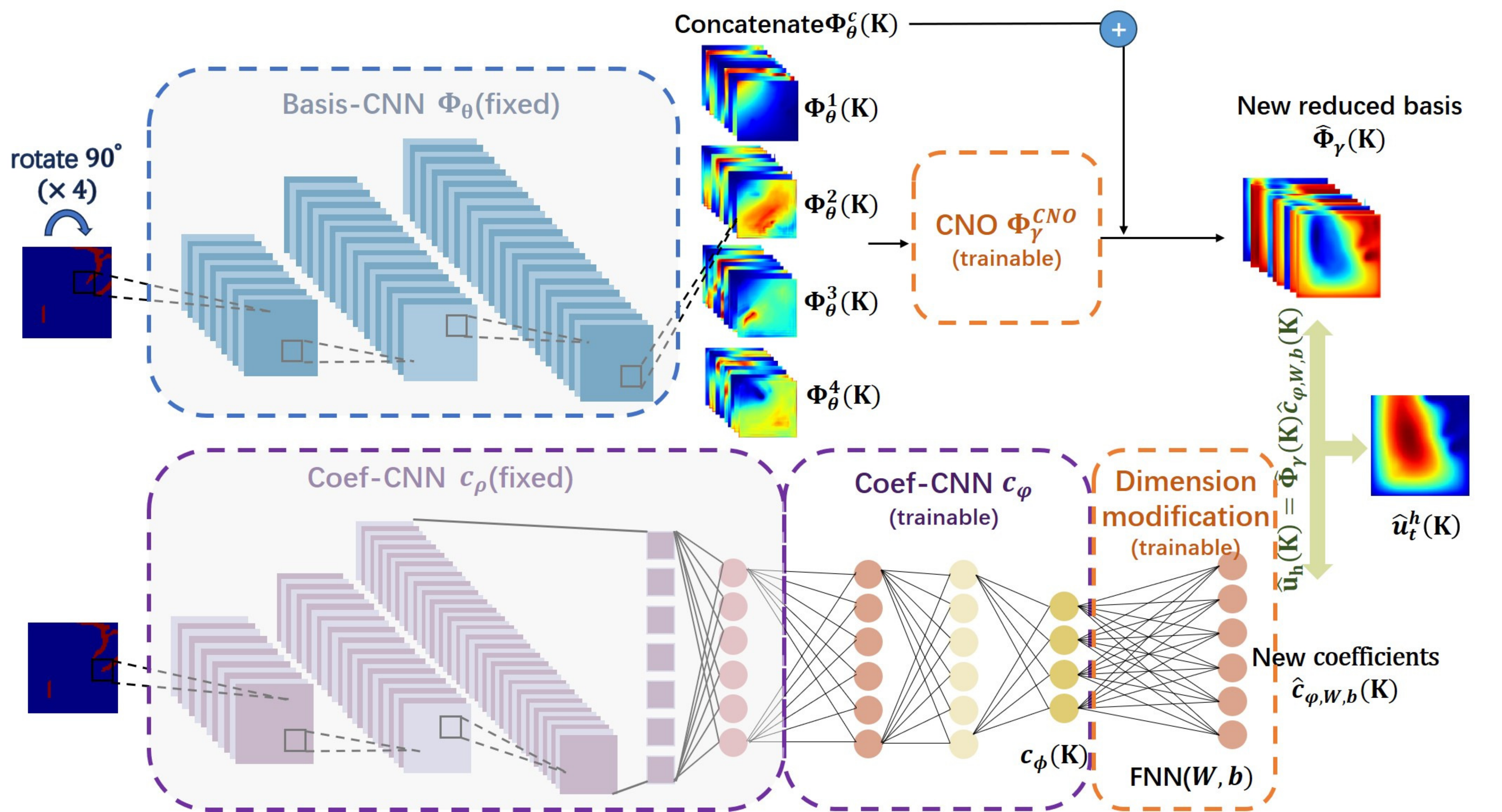}
	\caption{The two CNNs of MITL}	
	\label{sec4_fig:TL-CNNbasedROM}
\end{figure}

Based on the CNN-based ROM architecture, lightweight modifications are introduced to the network in this section to improve its adaptability to the target task. The new architecture also consists of two CNNs: TL-Basis  CNNs and TL-Coef CNNs (see Figure \ref{sec4_fig:TL-CNNbasedROM}). During the transfer learning process, the two CNNs are trained jointly.  TL-Basis CNNs are responsible for transferring and refining the reduced basis learned by the Basis CNNs from both data and governing equation. At the same time, TL-Coef CNNs retain the multiscale features learned by the Coef CNNs from the input data, while modifying only the high-level network architecture, thereby improving the adaptability of the overall network to the target task. The transfer strategy described above will be described in detail in the following subsections.

\subsubsection{Basis CNNs of MITL (TL-BasisCNNs)}
In Section \ref{ssec:MsFEM}, we note that the multiscale basis functions constructed by MsFEM are transferable to problems with different source terms and boundary conditions. Owing to this transferability, the reduced basis functions learned by the Basis CNNs from these global MsFEM basis functions are also capable of capturing transferable multiscale features. However, the four boundary conditions in source task (\ref{sec4_eq:source problem2}) originate from the four vertices of the computational domain and propagate outward, with their values decreasing linearly from 1 to 0. Consequently, each global basis function only captures a portion of the multiscale features. Therefore, for each input, it is necessary to fuse the features learned from the four global basis functions in order to obtain a complete feature representation of the output response. Sepecifically, The mapping that rotates a matrix by $\pi/2$ radians clockwise is defined as follows
\begin{equation}
	\begin{aligned}
		&\mathcal{R}_{\pi/2}:\mathbb{R}^{N_h\times N_h}\rightarrow\mathbb{R}^{N_h\times N_h},\\
		&A^{'} = \mathcal{R}_{\pi/2}(A),\quad A^{'} = [a_{j,N_h-i+1}],\quad A = [a_{ij}], \quad i=1,..,N_h,\quad j=1,\cdots,N_h.
	\end{aligned}
	\nonumber
\end{equation}
Similarly, we can define the mapping that rotates a matrix by $\pi/2$ radians anticlockwise by
\begin{equation}
	\begin{aligned}
		&\mathcal{R}_{-\pi/2}:\mathbb{R}^{N_h\times N_h}\rightarrow\mathbb{R}^{N_h\times N_h},\\
		&A^{'} = \mathcal{R}_{-\pi/2}(A),\quad A^{'} = [a_{N_h-j+1,i}],\quad A = [a_{ij}],  \quad i=1,..,N_h,\quad j=1,\cdots,N_h.
	\end{aligned}
	\nonumber
\end{equation}
Then, For a random input sample $K$, rotating it three times and passing all four versions through the trained Basis CNNs produces four sets of reduced-order basis functions, namely,
\begin{equation}
	\begin{aligned}
		&\Phi_{\theta}^{1}(K) := \Phi_{\theta}(K),\\
		&\Phi_{\theta}^{2}(K) := \mathcal{R}_{-\pi/2}(\Phi_{\theta}(\mathcal{R}_{\pi/2}(K))),\\
		&\Phi_{\theta}^{3}(K) := \mathcal{R}_{-\pi/2}\circ\mathcal{R}_{-\pi/2}(\Phi_{\theta}(\mathcal{R}_{\pi/2}\circ\mathcal{R}_{\pi/2}(K))),\\
		&\Phi_{\theta}^{4}(K) := \mathcal{R}_{\pi/2}\circ\mathcal{R}_{\pi/2}\circ\mathcal{R}_{\pi/2}(\Phi_{\theta}(\mathcal{R}_{\pi/2}\circ\mathcal{R}_{\pi/2}\circ\mathcal{R}_{\pi/2}(K)))
	\end{aligned}
	\nonumber
\end{equation}
By finally concatenating the above reduced basis, we can obtain a set of reduced-order basis $\Phi_{\theta}^{c}(K)$ from source task. The learning of these basis is constrained by the operator $-\nabla\cdot\Big(\kappa(x,\xi)\nabla \cdot\Big)$, where the operator itself depends only on the random input $\kappa(x,\xi)$. Thus, we can obtain the total multiscale reduced basis $\Phi_{\theta}^{c}(K)$, i.e.,
\begin{equation}
	\Phi_{\theta}^{c}(K) = [\Phi_{\theta}^{1}(K),\Phi_{\theta}^{2}(K),\Phi_{\theta}^{3}(K),\Phi_{\theta}^{4}(K)].
	\nonumber
\end{equation}
Compared with MsFEM, for a new sample $K$, the proposed method does not need to solve the cell problem (\ref{sec4_eq:source problem1}) for constructing local multiscale basis functions. Instead, reduced basis that encode multiscale features can be directly obtained by simply utilizing the trained Basis CNNs.

Since we also consider problems with different differential operator, the basis obtained by source task are not directly transferable to the target task. To this end, a residual structure is introduced to correct the reduced basis functions, namely,
\begin{equation}
	\hat{\Phi}_{\gamma}^{c}(K) = \Phi_{\theta}^{c}(K)+\Phi_{\gamma}^{\text{CNO}}\Big(\Phi_{\theta}^{c}(K)\Big),
	\nonumber
\end{equation}
where the residual is learned by convolutional neural operator (CNO) \cite{sec4_CNO}, $\gamma$ are trainable neural network parameters. There are two reasons for choosing CNO. Firstly, the core architecture of the CNO is also based on convolutional layers, which aligns with that of CNN-based ROM. This can ensure better compatibility. Second, it has been widely shown in the literature that the CNO can effectively capture multiscale features \cite{sec4_CNO}. 

In addition, the skip connections in the residual structure allow the original input information to be directly propagated to subsequent layers. This helps preserve the multiscale features learned from the source task, while improving adaptability to the target task.
\begin{rem}
	In this paper, we take MsFEM basis as an example. The rotation radian can take some random values to capture the  patterns of the input  field $\kappa(x,\xi)$.
\end{rem}
\subsubsection{Coefficient CNNs of MITL (TL-Coef CNNs)}
In TL-CoefCNNs, we employ paramter transfer strategy. Specifically, the CNN components and the first feedforward neural network (FNN) layer are frozen, while only the last two FNN layers are fine-tuned to adapt to the target task. Moreover, the number of reduced basis increases from $N$ to $4N$, additional linear layer is required to adjust output dimension, called dimension modification layer, i.e.,
\begin{equation}
	\hat{c}_{\varphi,W,b}(K) = W(c_{\varphi}\circ c_{\rho}(K))+b,
	\nonumber
\end{equation}
where $c_{\varphi}$ denotes the last two forward feedback neural networks of Coef CNN and $c_{\rho}$ represents convolutional parts. In many image processing tasks, shallow layers are considered to capture general features of images (e.g., edges, corners, and textures), while deeper layers are responsible for extracting high-level semantic information \cite{sec4_CNN}. Since the input samples and system responses can also be regarded as images, the parameters of $c_{\rho}$ are frozen in target tasks while $\varphi$ is set to be trainable trainable. $W\in \mathbb{R}^{4N \times N}$ and $b\in \mathbb{R}^{4N}$ are trainable parameters in the dimension modification layer, which used to change the dimension of coefficients from $N$ to $4N$.

\subsubsection{Loss function}
Based on the previous two subsections, we can construct a surrogates for target task, i.e.,
\begin{equation}
	\hat{u}_t^{h}(K_i) = \hat{\Phi}_{\gamma}(K_i)\hat{c}_{\varphi,W,b}(K_i),
	\nonumber
\end{equation}
where $\gamma,\varphi,W,b$ are trainable paramters of neural networks. Given the dataset 
\begin{equation*}
	\mathcal{D}_t = \Big\{K_i,u_t^h(K_i)\Big\}_{i=1}^{N_t},
\end{equation*} 
the following optimization problem is solved
\begin{equation}
	\arg\min_{\gamma,\varphi,W,b}\frac{1}{N_t}\sum_{i=1}^{N_t}\Vert u_t^{h}-\hat{u}_t^{h}\Vert_2^2,
	\nonumber
\end{equation}
where $\Vert \cdot\Vert_2$ is matrix $2-$norm.
\begin{rem}
	\textbf{MITL can also be used to unsteady problems.} We assume that the multiscale random input field $\kappa(x,\xi)$ is only depends on the spatial variable $x$. Take the following parabolic problems as an example,
	\begin{equation*}
		\left\{
		\begin{aligned}
			&\frac{\partial u(x,t)}{\partial t}=\nabla \cdot \Big( \kappa(x,\xi)\nabla u(x,t)\Big)+f(x),\quad x \in \Omega,\\
			&\mathcal{D}u(x,t)=b(x),\quad x \in \partial{\Omega},\\
			&u(x,0) = u_0(x),\quad x\in \Omega,
		\end{aligned}
		\right.
		\label{sec2_eq:parabolic}
	\end{equation*}
	
	Therefore, only TL-Coef CNN is required to be modified to learn a time-dependent cofficients for the target tasks. Specifically, a shallow feedforward network $c_{\alpha}$ is introduced to encode time. The resulting temporal encoding is then added to the spatial features and is passed to the subsequent network (see Figure \ref{sec4_fig:TL-CoefCNNtime}). Such process can be written as
	\begin{equation}
		\hat{c}_{\varphi,W,b,\alpha}(K,t) = W\Big(c_{\varphi}\circ\big(c_{\rho}(K)+c_{\alpha}(t)\big)\Big)+b.
		\nonumber
	\end{equation}
	\begin{figure}[htp]
		\centering
		\includegraphics[scale=0.26]{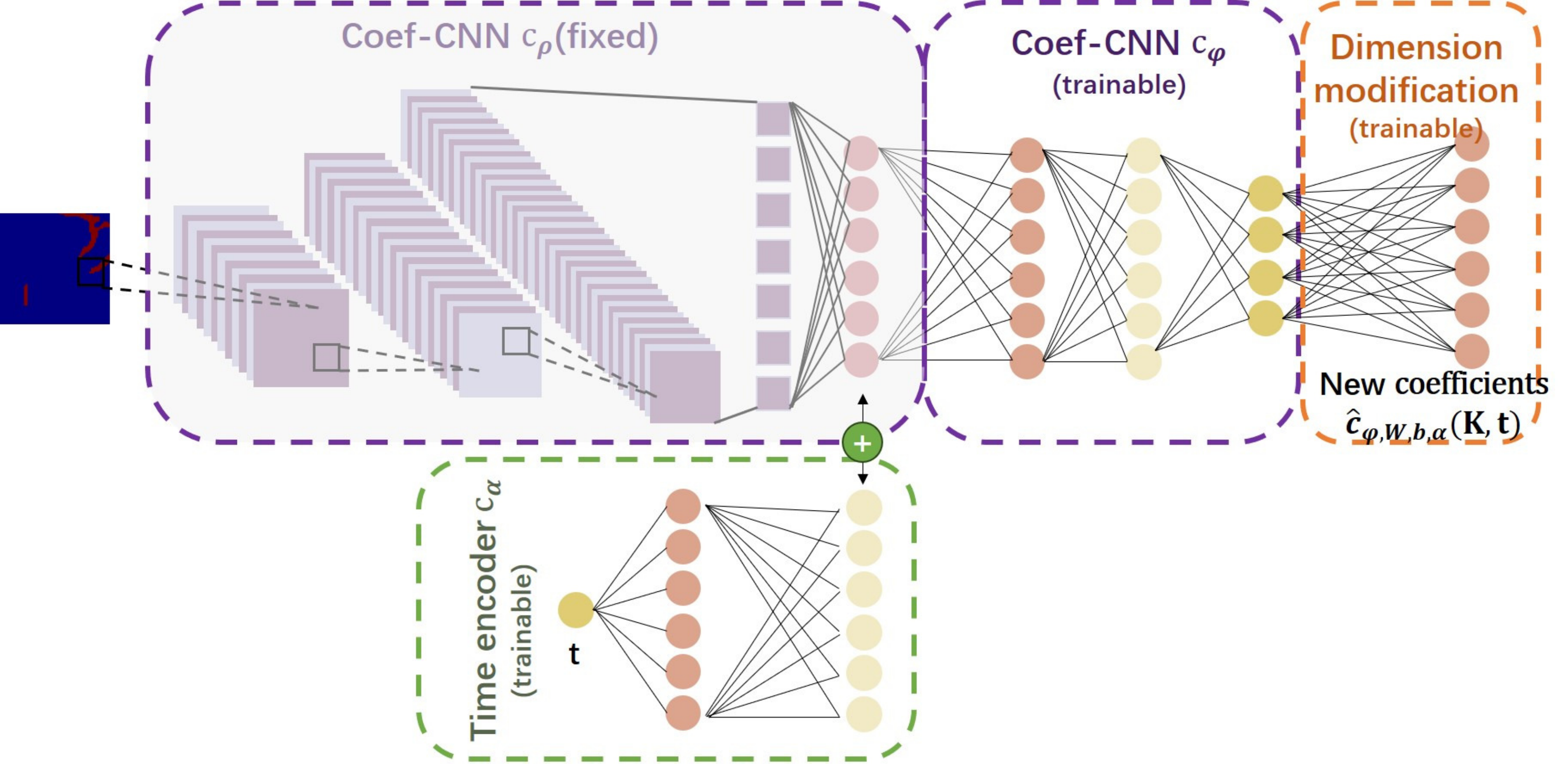}
		\caption{TL-Coef  CNNs for unsteady problems}	
		\label{sec4_fig:TL-CoefCNNtime}
	\end{figure}
	Thus, TL-Coef CNN is capable of predicting the solution at any time $t\in(0,T]$,
	\begin{equation}
		\hat{u}_t^h(K,t) = \hat{\Phi}_{\gamma}(K)\hat{c}_{\varphi,W,b,\alpha}(K,t).
	\end{equation}
\end{rem}

\subsubsection{Nonlinear nature of the mapping from multiscale basis to fine-scale solutions}
The MsFEM origins from homogenization theory, which illustrate that the mapping from multiscale basis to fine-scale solutions is naturally nonlinear. Speciffically, consider the following highly oscillating periodic operator with macro-micro ratio $\epsilon \ll 1$,
\begin{equation}
	-\nabla \cdot \Big(\kappa(\frac{x}{\epsilon}\Big)\nabla u_{\epsilon}(x)\Big) = f(x), x\in \Omega,
	\nonumber
\end{equation}
where $\kappa(\frac{x}{\epsilon})$ is $y-$periodic with $y = x/ \epsilon$. By solving cell problems \cite{sec4:homogenization}, we can obtain homogenized coeffcients $\kappa^{*}$ and periodic basis $\Phi_{\text{ms}} = (\phi_1,\cdots,\phi_d)$ where $d$ is the dimension of spatial domain. Then, we can obtain the coarse solution $u^{*}$, which satisfy the corresponding coarse model,
\begin{equation}
	-\nabla \cdot \Big(\kappa^{*}(x)\nabla u^{*}(x)\Big) = f(x), x\in \Omega.
	\nonumber
\end{equation}
According to asymptotic expansion, multiscale solutions $u_{\epsilon}(x)$ can be written as \cite{sec4:homogenization}
\begin{equation}
	u_{\epsilon}(x) = u^{*}(x)+\epsilon \nabla u^{*}(x)\cdot \Phi_{\text{ms}}(x/\epsilon) + O(\epsilon^2) = u^{*}\Big(x+\epsilon \Phi_{\text{ms}}(x/\epsilon)\Big) + O(\epsilon^2).
	\nonumber
\end{equation}
Therefore, $u_{\epsilon}$ is a nonlinear function of basis $\Phi_{\text{ms}}$. The design of MITL is guided by this idea, that is, 
\begin{equation}
	\hat{u}_t^{h} = \hat{\Phi}_{\gamma}\hat{c}_{\varphi,W,b} = (I+\Phi_{\gamma}^{\text{CNO}})(\Phi_{\theta}^c)\hat{c}_{\varphi,W,b}
	 = \Big(c\circ(I+\Phi_{\gamma}^{\text{CNO}})\Big)(\Phi_{\theta}^c),
\end{equation} 
where $c:\mathbb{R}^{N_h^2\times4N}\rightarrow \mathbb{R}^{4N\times 1}$ and is defined by  $c(\Phi):= \Phi \hat{c}_{\varphi,W,b}$. Thus, the prediction  soluton $\hat{u}_t^{h}$ is  nonlinear with respect to  MsFEM reduced basis $\Phi_{\theta}^c$  through the nonlinear mapping $c\circ(I+\Phi_{\gamma}^{\text{CNO}})$.

\section{Numerical experiment}
\label{sec5:numerical experiment}
In this section, the numerical results for various multiscale problems with random inputs
are presented to show the accuracy and efficiency of MITL.

To illustrate numerical results quantitatively, we define the relative $L_2$ error $\epsilon_{\text{test}}$ as follows,
\begin{equation}
	\epsilon_{\text{test}} = \frac{1}{D}\sum_{i=1}^D\frac{\Vert u_i-\hat{u}_i \Vert_2}{\Vert u_i \Vert_2},
	\nonumber
\end{equation} 
where $u_i$ denotes the results of numerical simulations and $\hat{u}_i$ is the approximation solution obtained by MITL. And $D$ represents the size of test dataset. The meaning of them varies across different tasks. For source tasks, $u_i = \Phi_{\text{ms}}(K_i),\quad \hat{u}_i = \hat{\Phi}_{\text{ms}}(K_i),\quad D = N_s^{test}$. For target tasks, two cases are distinguished: for steady problems, $u_i = u^{h}(K_i),\quad \hat{u}_i = \hat{u}^{h}(K_i),\quad D = N_t^{test}$; for unsteady problems, $u_i = u_t^{h}(K_k,t),\quad \hat{u}_i = \hat{u}_t^{h}(K_k,t),\quad D = N_t^{test}*T_t$.

\subsection{The construction of datasets}
\label{ssec:construction of dataset}
In this section, we first introduce two types of random input fields: hydraulic conductivity field of an aquifer with channels and permeability of fractured rock. They are both binary-valued, representing the foreground and background media, respectively. The geometry and position of foreground media are stochastic. 

\subsubsection{Hydraulic conductivity field of  channelized aquifers}
\label{ssec:channelized aquifers}
We first consider a relatively idealized random field, representing the hydraulic conductivity of a channelized aquifer. The foreground corresponds to the channels and the background to the matrix material, with their hydraulic conductivities set to 1000 and 1, respectively. The primary complexity of this random field lies in the varying branching patterns of the channels across different samples, while the overall orientation and width of the channels remain relatively fixed. By using a trained generative model \cite{sec5_K_groundwater}, 6024 samples of this random field are obtained, of which 5024 are used for training the source and target tasks, and the remaining 1000 are reserved for evaluating the generalization capability of the model. 
\begin{figure}[htp]
	\centering
	\includegraphics[scale=0.48]{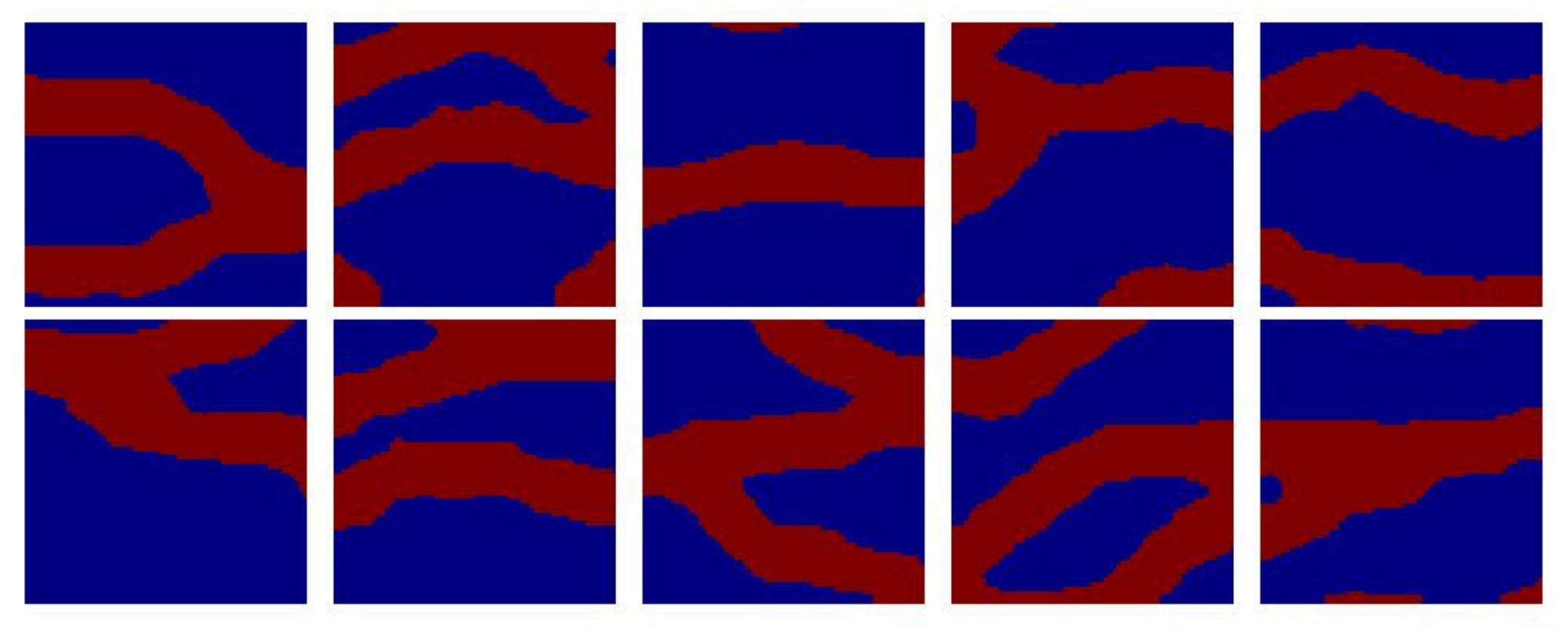}
	\caption{Eight samples of the hydraulic conductivity of channelized aquifers}	
	\label{sec5_fig:groundwater_K_sample}
\end{figure}
\subsubsection{Permeability of fractured rock}
\label{ssec:fractured rock}
\begin{figure}[htp]
	\centering
	\includegraphics[scale=0.32]{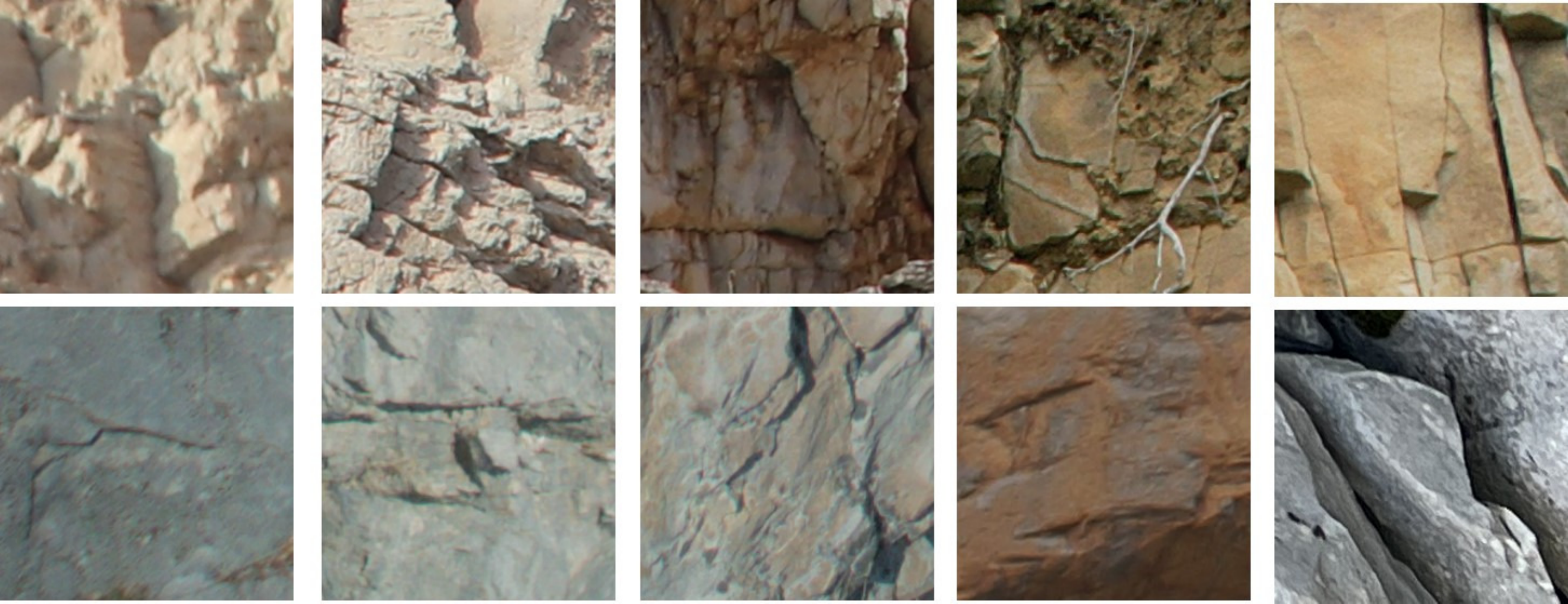}
	\caption{The real-world photos of fractured rocks  \cite{sec5_K_fracture}}	
	\label{sec5_fig:fracture_K_sample_real}
\end{figure}
\begin{figure}[htp]
	\centering
	\includegraphics[scale=0.48]{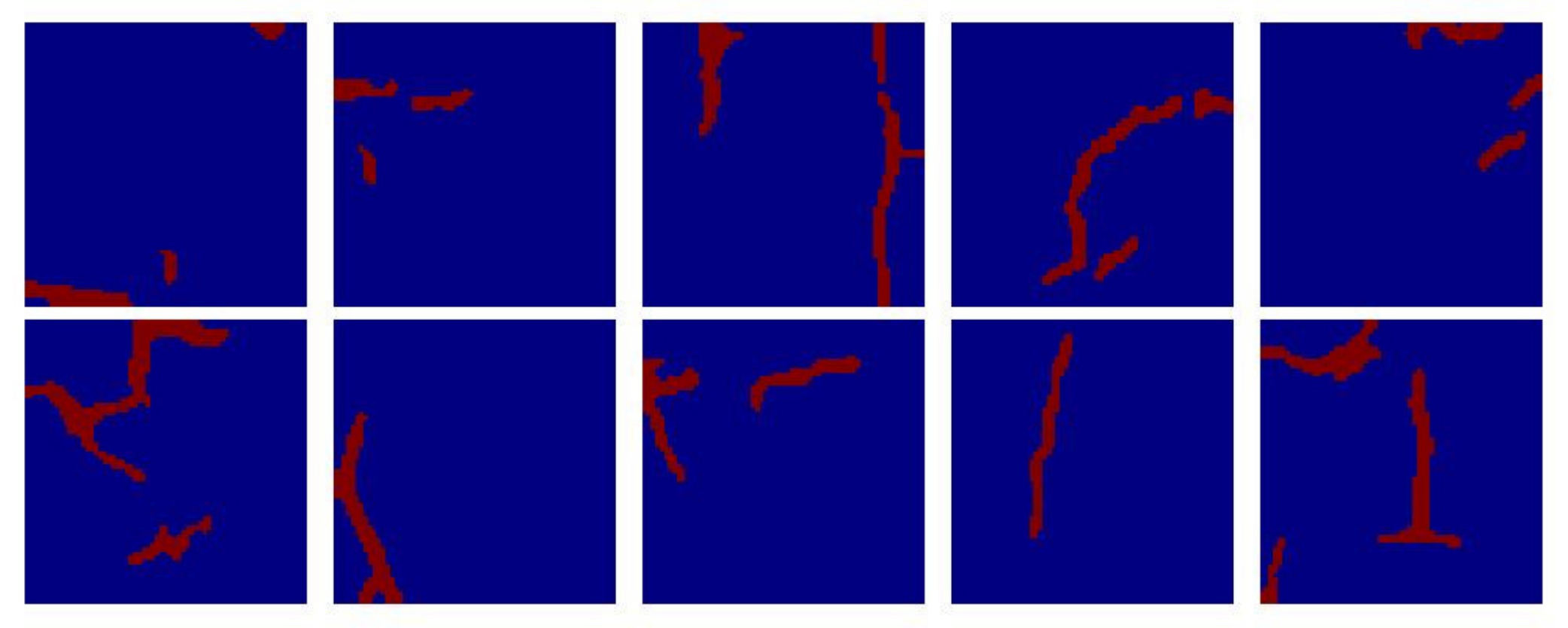}
	\caption{Eight samples of the permeability of fractured rocks}	
	\label{sec5_fig:fracture_K_sample}
\end{figure}

The random fields considered in this subsection are sourced from the GeoCrack dataset \cite{sec5_K_fracture}, which represent the permeability of fractured rocks. The rock samples in this dataset were collected from Europe (Greece and Italy) and the Middle East (Oman and the United Arab Emirates), with the fracture traces manually annotated. The real fracture geometries (including width, branching patterns, orientation, curvature, density and so on) exhibit considerably greater complexity (see Figure \ref{sec5_fig:fracture_K_sample_real}).
To control sample variability, rock images are selected based on both fracture morphology similarity and sampling location. The selected images are binarized and localized, yielding subimages of resolution $65 \times 65$ in which the fracture area constitutes $3\%-10\%$ of the total area. Finally, 6024 samples were obtained, of which 5024 were assigned to the training set and the remainder to the test set. For simplicity, the fracture permeability is taken as 1000 and the rock matrix permeability as 1. Eight processed samples are shown in Figure \ref{sec5_fig:fracture_K_sample}.

In order to construct training and test datasets, bilinear finite element methods in rectangular elements are be used to solve source task (\ref{sec4_eq:source problem2}) and target rask (\ref{sec2_eq:2ord linear})-(\ref{sec2_eq:parabolic}). In the finite element formulation, the random input field $\kappa$ is sampled at the single-point Gaussian quadrature points, while the training process requires its values at the nodal points of spatial mesh $\mathcal{H}$. To address this issue, the values of the random field on $\mathcal{H}$ are interpolated onto the quadrature points using the nearest-neighbor interpolation scheme. Specifically, for any matrix $K = [k_{ij}]\in \mathbb{R}^{N_h\times N_h}$, the interpolated matrix $\hat{K}=[\hat{k}_{ij}] \in \mathbb{R}^{(N_h-1)\times (N_h-1)}$ is obtained through the mapping
\begin{equation}
	\hat{k}_{ij} = k_{[iN_h/(N_h-1)],[jN_h/(N_h-1)]}.
	\nonumber
\end{equation}
This interpolation method is more suitable for binary-valued matrices. 
\subsection{Source tasks}
\begin{figure}[htp]
	\centering
	\begin{subfigure}[b]{1.\textwidth}
		\centering
		\includegraphics[width=\textwidth]{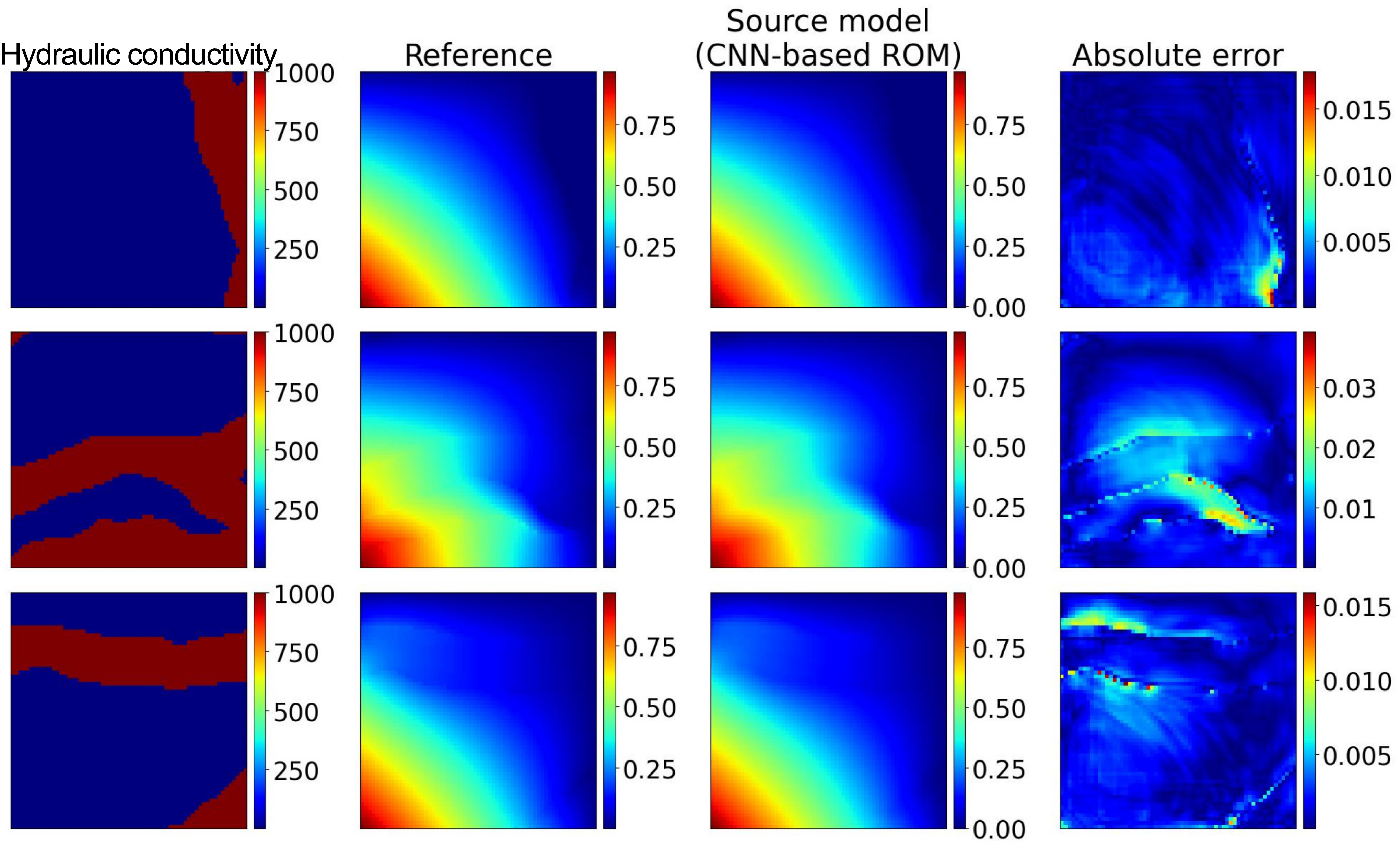}
		\caption{Channelized aquifers}
	\end{subfigure}
	\hfill
	\begin{subfigure}[b]{1,\textwidth}
		\centering
		\includegraphics[width=\textwidth]{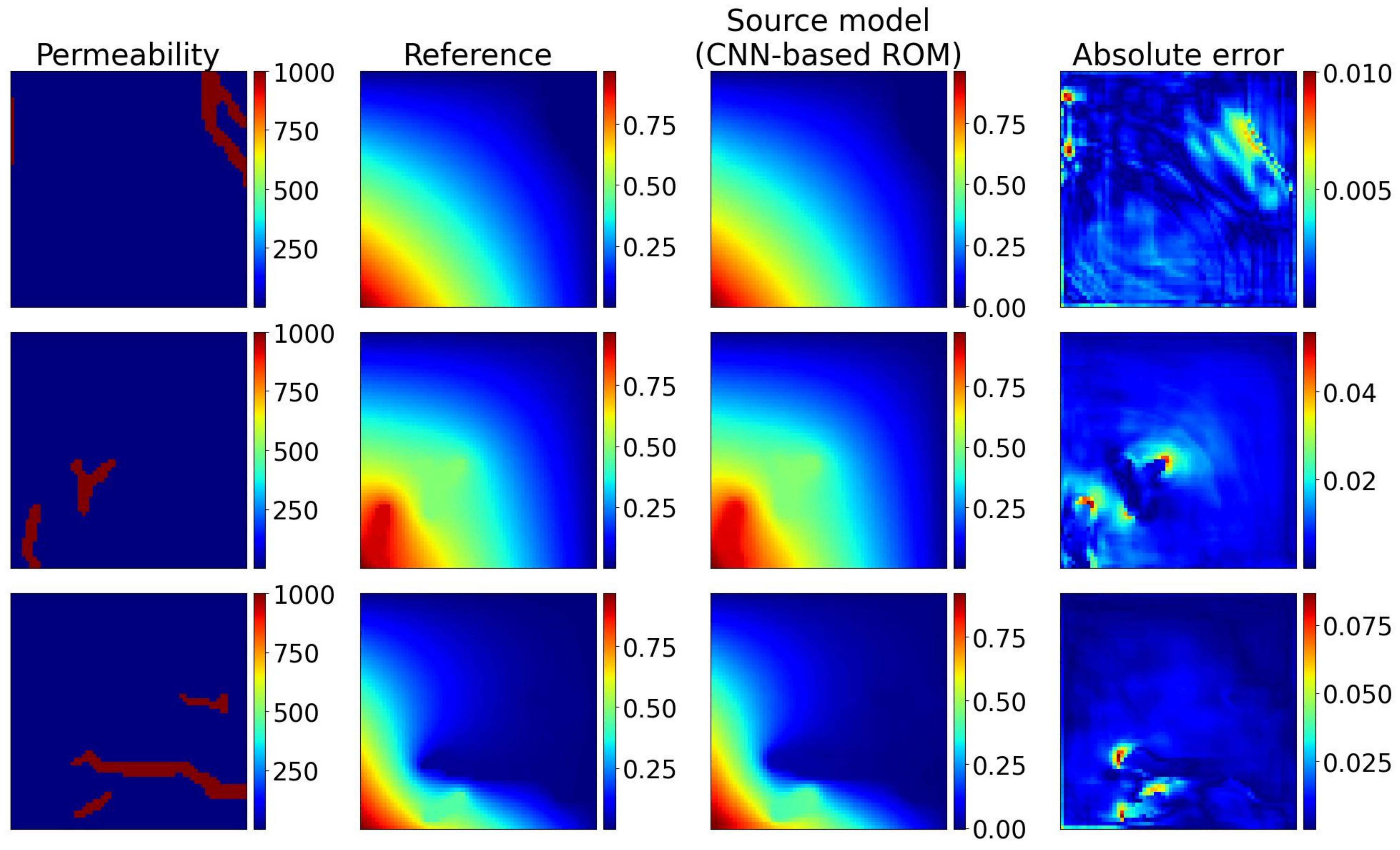}
		\caption{Fractured rocks}
	\end{subfigure}
	\caption{The prediction performance of the source model}
	\label{sec5_fig:source model}
\end{figure}
In MITL, source tasks (\ref{sec4_eq:source problem2}) are first approximated by CNN-based ROM to obtain transferable features. Rotating each random field sample defined in Section \ref{ssec:construction of dataset} by $90^\circ$
, $180^\circ$ and $270^\circ$ yields a total of 24,000 samples, among which 20000 samples are splitted to training dataset and the rest ones for test dataset. The corresponding source tasks are solved using the finite element method on the mesh $\mathcal{H}$ of a $65\times 65$ grid. Then, we obtain the source-task dataset $\mathcal{D}_s$. The architecture of CNN-based ROM used for the source tasks is listed in Table \ref{sec5_table:architecture of CNNbasedROM}.
\begin{table}[H]
	\centering
	\setlength{\abovecaptionskip}{0cm}
	\setlength{\belowcaptionskip}{0.2cm}
	\caption{The architecture of CNN-based ROM used for the source tasks (\ref{sec4_eq:source problem2})}
	\label{sec5_table:architecture of CNNbasedROM}
	\resizebox{\textwidth}{!}{
		\begin{tabular}{c|c|c||c|c|c}
			\hline
			\multicolumn{3}{c||}{\textbf{Basis CNNs}} & \multicolumn{3}{c}{\textbf{Coef CNNs}} \\[2ex]
			\hline
			Layer & Resolution $H \times W$ & Parameter numbers & Layer & Resolution $H \times W$ & Parameter numbers \\
			\hline
			Input & $63\times 63$ & - & Input & $63\times 63$ & - \\
			Conv$_\text{p}$ (25$\times$25,32,1) & $63\times 63$ & 20064 & Conv (23$\times$23,32,1) & $63\times 63$ & 16992 \\
			Conv$_\text{p}$ (25$\times$25,32,1) & $63\times 63$ & 640064 & Conv (23$\times$23,32,2) & $32\times 32$ & 541760 \\
			Conv$_\text{p}$ (25$\times$25,64,1) & $63\times 63$ & 1280128 & Conv (23$\times$23,64,1) & $32\times 32$ & 1083520 \\
			Conv$_\text{p}$ (25$\times$25,64,1) & $63\times 63$ & 2560128 & Conv (23$\times$23,64,2) & $16\times 16$ & 2166912 \\
			Conv$_\text{p}$ (25$\times$25,128,1) & $63\times 63$ & 5120256 & Conv (23$\times$23,128,1) & $16\times 16$ & 4333824 \\
			Conv$_\text{p}$ (25$\times$25,128,1) & $63\times 63$ & 10240256 & Conv (23$\times$23,128,2) & $8\times 8$ & 8667392 \\
			Conv (25$\times$25,1,1) & $63\times 63$ & 1280 & FC (300) & $300$ & 2457900 \\
			Reshape & $3969\times 10$ & - & FC (300) & $300$ & 90300 \\
			& & & FC (10) & $10$ & 3010 \\
			\hline
			\multicolumn{1}{c}{\textbf{Total number of parameters}} & \multicolumn{2}{c}{} & \multicolumn{3}{r}{39223786}\\
			\hline
		\end{tabular}
	}
\end{table}
The Basis CNNs and Coef CNNs are trained sequencely by using the methods in our previous work \cite{sec3_CNNbasedROM}. Then, we can obtain the source models (\ref{sec4_eq:source model}). As shown in Table \ref{sec5_table:source tasks}, CNN-based ROM can provide accurate surrogates for source tasks. Since the source tasks with permeability of fractured rocks are more complex, the relative $L_2$ test error is larger than that of channelized aquifers using the same neural network architecture.
\begin{table}[H]
	\centering  
	\setlength{\abovecaptionskip}{0cm}
	\setlength{\belowcaptionskip}{0.2cm}
	\caption{The relative $L_2$ test error of CNN-based ROM used for source tasks}  
	\label{sec5_table:source tasks}  
	\scalebox{1.}{
		\begin{tabular}{ccc}
			\toprule  
			& channelized aquifers & fractured rocks   \\  
			\midrule
			$\epsilon_{\text{test}}$ & 1.52\%&  2.72\% \\			
			\bottomrule
		\end{tabular}
	}
\end{table}
Figure \ref{sec5_fig:source model} provides a more intuitive illustration of the prediction performance of the source model on three test samples. 

\begin{figure}[htp]
	\center
	\includegraphics[scale=0.30]{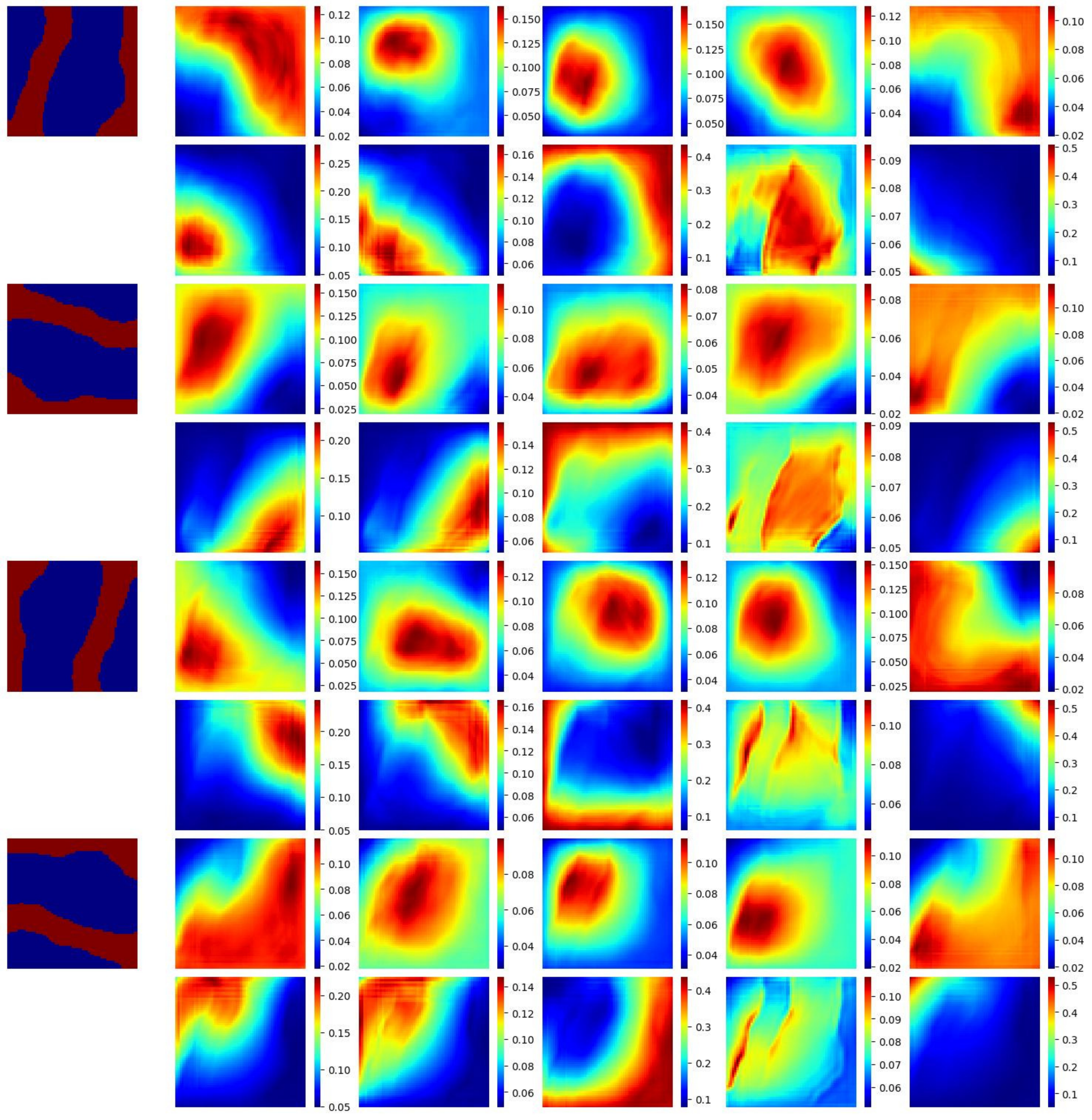}
	\caption{Channelized aquifers: reduced basis learned by source model}	
	\label{sec5_fig:reduced_order_basis}
\end{figure}

Since MITL needs to transfer the reduced basis learned by the Basis CNNs, we plot 40 reduced basis learned by the CNN-based ROM for one input sample $K$ of channelized aquifers (see Figure \ref{sec5_fig:reduced_order_basis}). The figure shows that the Basis CNNs capture the channel morphology in the learned basis functions. The basis functions with larger magnitudes focus on the multiscale features emanating from the four corners, consistent with the MsFEM basis functions.

\subsection{Target tasks}
In this section, the multiscale features learned by the source model are transferred to a series of target tasks, which involves different source terms, boundary conditions, and even different differential operators. To realize efficient transfer, we propose MITL. The core strategy is to freeze the main architecture of the source model and fine-tune only a small number of specific layers to adapt to new tasks. 

In details, TL-Basis CNNs freeze all trainable parameters of Basis CNNs to inherit its capability of extracting multiscale features. Only the two-layer CNO, originally designed for learning the residual term, is set as trainable to capture the new features of target tasks.
This trainable portion comprises 183,344 parameters, accounting for merely $0.92\%$ of the total parameters of Basis CNNs in the source model. In TL-Coef CNNs, the backbone network responsible for feature extraction is frozen, and only the last two feedforward layers together with the temporal encoding network and the dimension modification layer are trained. For steady and unsteady problems, the numbers of trainable parameters are 93,310 and 90,900 (plus 440 parameters from the dimension adaptation layer), accounting for only $0.48\%$ and $0.95\%$ of the total parameters of Coef CNNs in the source model, respectively. Overall, the trainable parameters of MITL framework amount to only 
$0.94\%$ of those of CNN-based ROM defined in Talbe \ref{sec5_table:architecture of CNNbasedROM}. Compared with the source model, the target model is considerably more lightweight. 
\subsubsection{Transfer to problems with different source terms}
\begin{figure}[htp]
	\centering
	\includegraphics[scale=0.23]{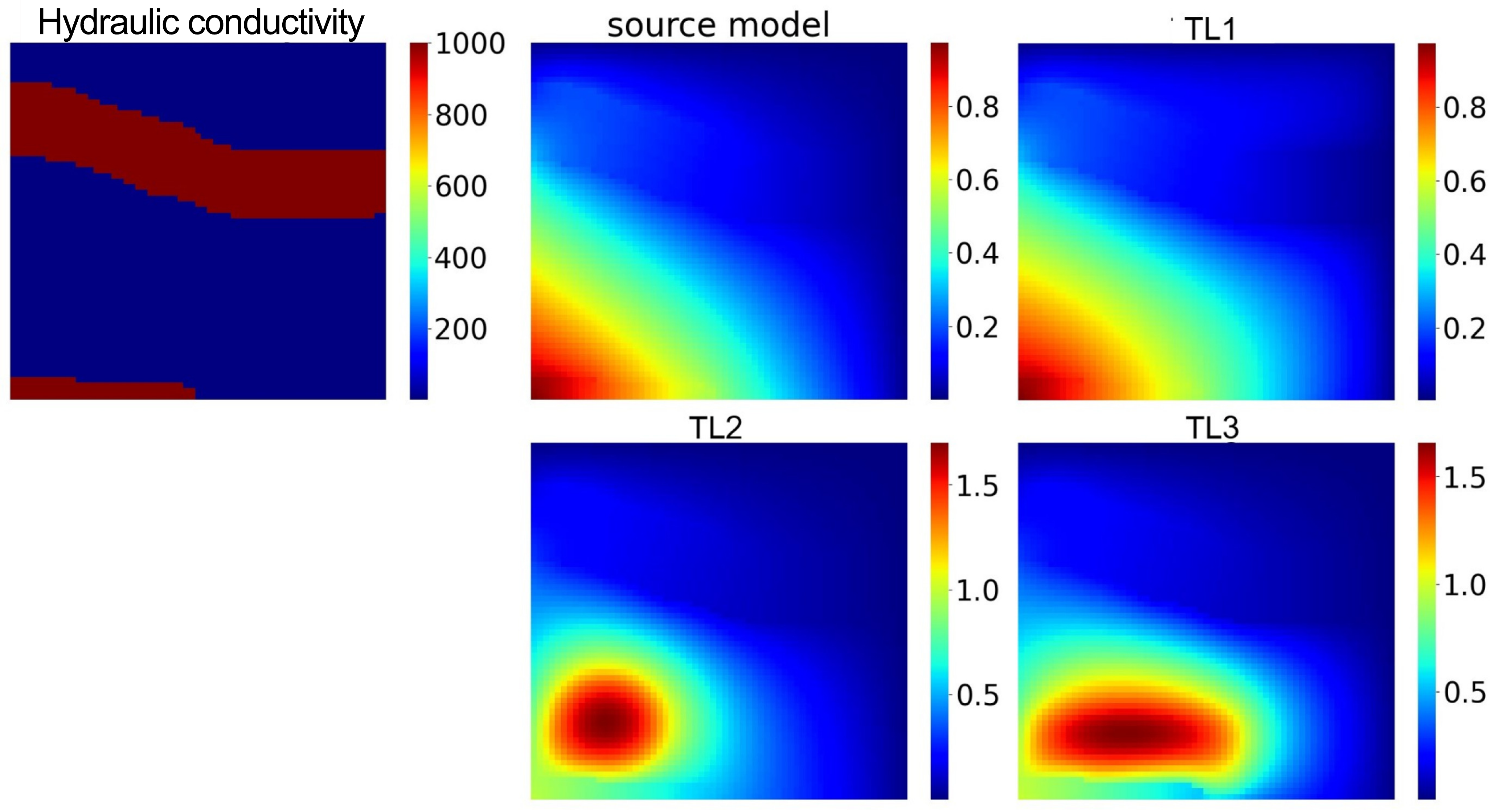}
	\caption{Channelized aquifers: solution samples of problems with three different source terms}	\label{sec5_fig:groundwater_differentsourceterm_sample}
\end{figure}

\begin{figure}[htp]
	\centering
	\includegraphics[scale=0.25]{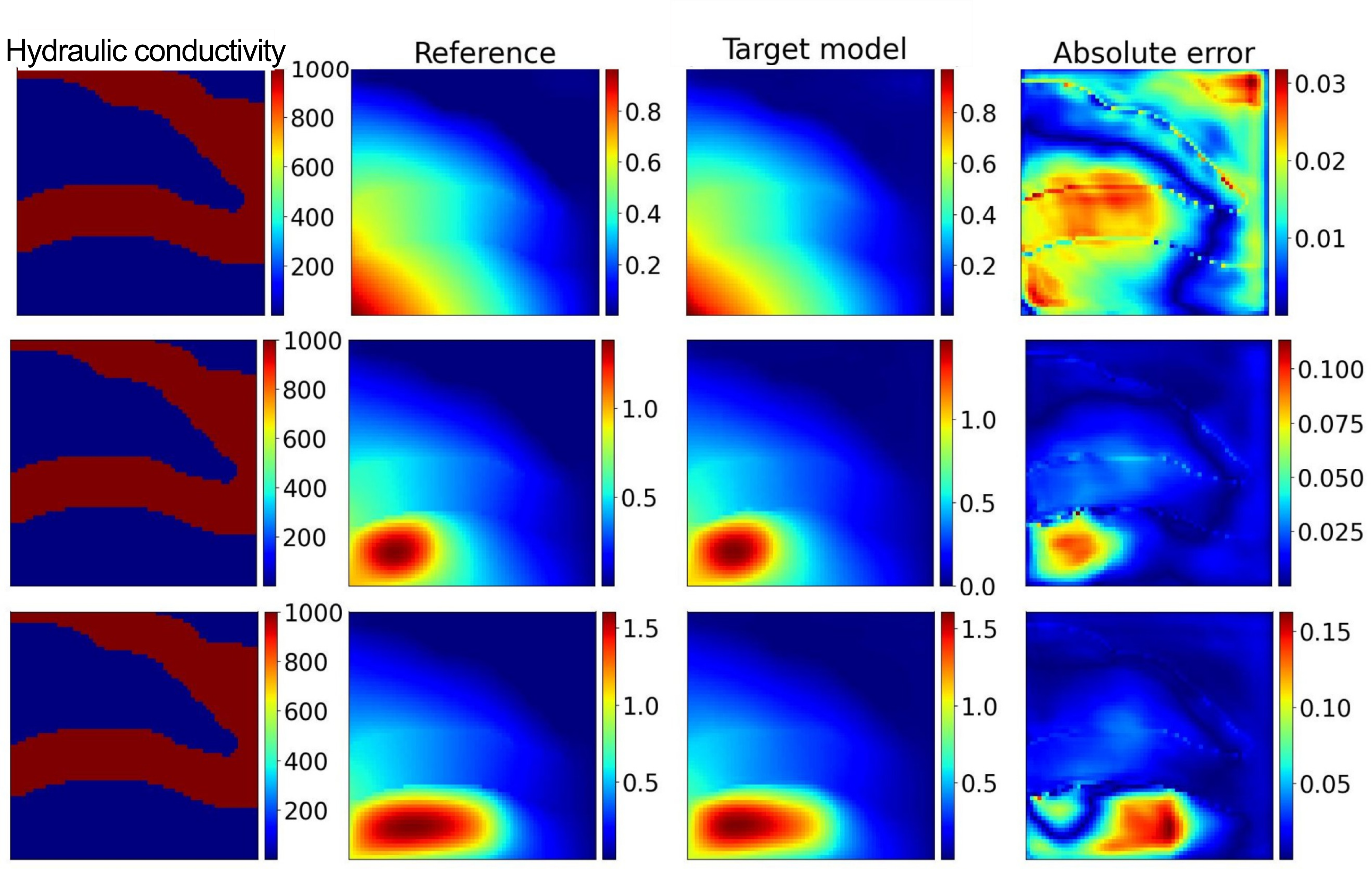}
	\caption{Channelized aquifers: MITL performance for problems with different source terms (From top to bottom row: TL1, TL2, TL3, the size of training dataset is 32.)}	
	\label{sec5_fig:groundwater_differentsourceterm}
\end{figure}

We first consider problems with different source terms, i.e., keeping the differential operator and the boundary conditions in the source tasks (\ref{sec4_eq:source problem2}) unchanged, and varying only the source term 
$f(x),x = [x_1,x_2]\in \Omega$. The following three source terms are considered:
\begin{itemize}
	\item \textbf{[TL1]Exponential function}: $f(x) = \exp(x_1+x_2).$
	\item \textbf{[TL2]Indicator function having a circle support}:
	\begin{equation}
		f(x)=\left\{
		\begin{aligned}
			&100,\quad x\in \{x|(x_1-0.2)^2+(x_2-0.2)^2\le 0.15^2\},\\
			&0,\quad  \text{ortherwise}.
		\end{aligned}
		\right.
		\nonumber
	\end{equation}
	\item \textbf{[TL3]Indicator function having a rectangular support}:
	\begin{equation}
		f(x)=\left\{
		\begin{aligned}
			&100,\quad x\in [0,0.6]\times[0,0.25],\\
			&0,\quad  \text{ortherwise},
		\end{aligned}
		\right.
		\nonumber
	\end{equation}
\end{itemize}
where TL1 is continuous with respect to $x$, while the latter two are discontinuous. In this section, the random field defined in Section \ref{ssec:channelized aquifers} is considered. For the same random input sample $K$, the solutions corresponding to these three source terms are shown in Figure \ref{sec5_fig:groundwater_differentsourceterm_sample}. From the perspective of data distribution, TL1 exhibits the closest distribution to that of the source task.
The target model is trained with training sets of sizes 5024, 1024, 512, 256, 128, and 64, respectively. The relative $L_2$ error is evaluated on the same test set of size 1000, and the results are presented in Table \ref{sec5_table:groundwater_differentsourceterm}. An instrusive result is also shown in Figure \ref{sec5_fig:groundwater_differentsourceterm}.
\begin{table}[H]
	\centering  
	\setlength{\abovecaptionskip}{0cm}
	\setlength{\belowcaptionskip}{0.2cm}
	\caption{Channelized aquifers: MITL performance for problems with different source terms}  
	\label{sec5_table:groundwater_differentsourceterm}  
	\scalebox{1.}{
		\begin{tabular}{ccccccc}
			\toprule  
			& 5024 & 1024 & 512 & 256 & 128 & 64 \\  
			\midrule
			TL1 & 2.12\%& 2.76\%& 2.93\%& 3.25\%& 3.50\%& 4.07\%\\
			TL2 & 2.49\%& 2.84\%& 3.17\%& 3.96\%& 5.44\%& 5.81\%\\
			TL3 &2.45\% &2.65\%&3.83\%&3.86\%&4.89\%&5.87\%\\							
			\bottomrule
		\end{tabular}
	}
\end{table}
The above results demonstrate that even when the target dataset $\mathcal{D}_t$ contains only 64 samples (accounting for merely $0.32\%$ of the source dataset), The relative $L_2$ error of different problems remain within an acceptable range. This result highlights the significant advantage of transfer learning in dealing with small-sample scenarios.

\subsubsection{Transfer to problems with different boundary conditions}
In this section, we consider more complex random field defined in Section \ref{ssec:fractured rock} and problems with differenet boundary conditions, including inhomogeneous Dirichlet boundary condition and mixed Dirichlet–Neumann boundary conditions. Speciffically, the boundary operators $\mathcal{D}$ are defined as follows,
\begin{itemize}
	\item \textbf{[TL4] Inhomogeneous Dirichlet boundary condition}
	\begin{equation*}
		u(x) = \left\{
		\begin{aligned}
			&1,\quad x \in x\in \partial \Omega_D^1,\\
			&0,\quad x\in \partial \Omega_D^0,
		\end{aligned}
		\right.
	\end{equation*}
	where $\partial \Omega_D^1 = \{x:=(x_1,x_2)\in\partial  \Omega: x\in \big(\{0\}\times[0.25,0.75]\big)\bigcup\big([0.6,1] \\ \times[0.25,1]\big)\}, \partial \Omega_D^0 = \Omega \backslash \Omega_D^1$.
	\item \textbf{[TL5] Mixed Dirichlet–Neumann boundary conditions}
	\begin{equation}
		\left\{
		\begin{aligned}
			&u(x)=1,  \quad x \in \partial{\Omega}_D^1, \\
			&u(x)=0, \quad x \in \partial{\Omega}_D^0,\\
			&\frac{\partial{u}}{\partial{\vec{n}}}=0, \quad x \in \partial{\Omega}_N,
		\end{aligned}
		\right.
		\nonumber
	\end{equation}
	where $	\partial{\Omega}_N=\{x:=(x_1,x_2)\in\partial{\Omega}:x_2=1\}, \ \partial{\Omega}_D^1=\{x\in\partial{\Omega}:x_2=0\},\ \partial{\Omega}_D^0=\partial{\Omega}\backslash\partial{\Omega}_N\backslash \partial{\Omega}_D^1$.
\end{itemize}
\begin{figure}[htp]
	\centering
	\includegraphics[scale=0.24]{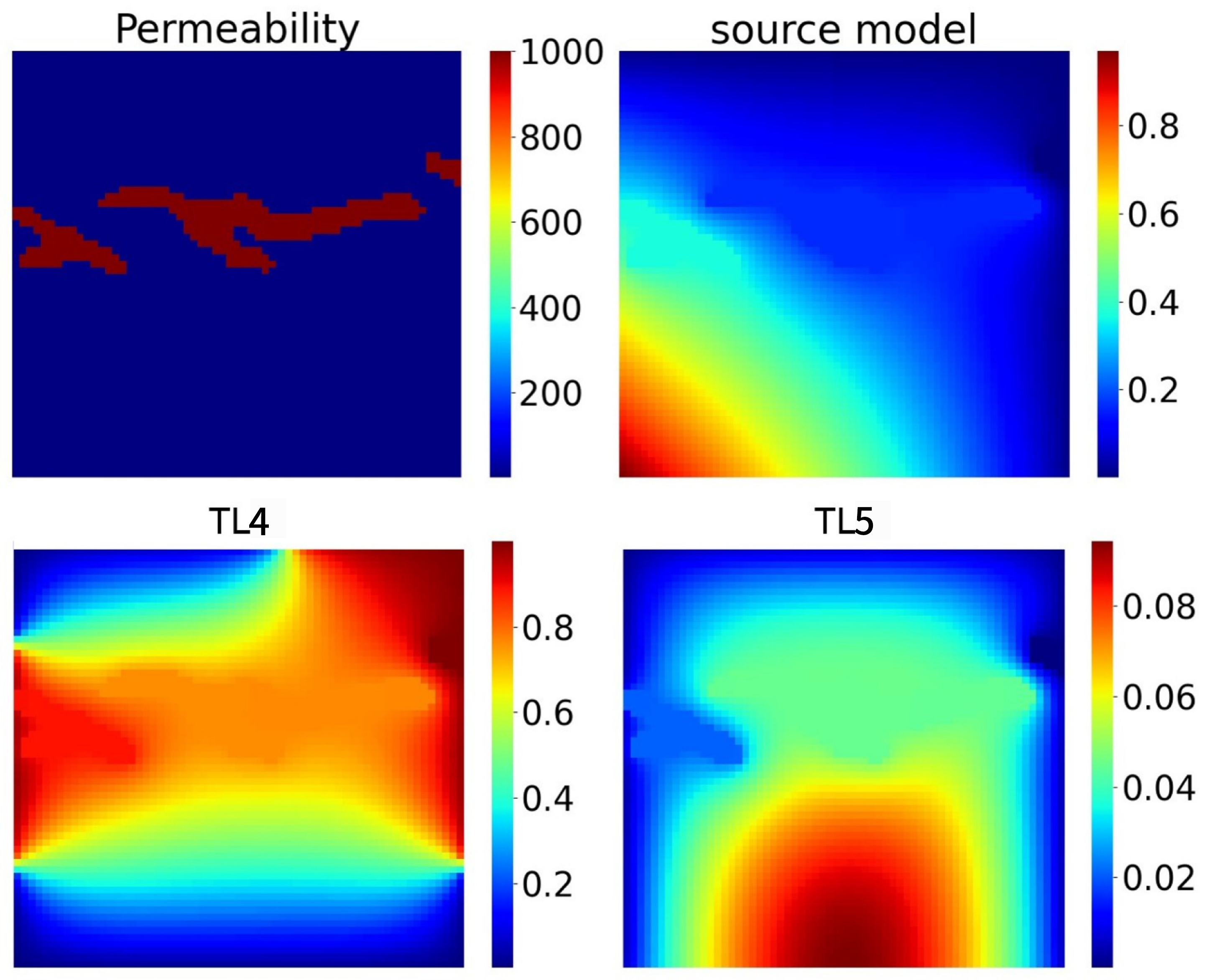}
	\caption{Fractured rocks: solution samples of problems with two different boundary conditions}	\label{sec5_fig:fracture_differentboundarycondition_sample}
\end{figure}

In addition, the source term is taken as $f(x)=1$. For the same input sample $K$, the solutions corresponding to these two boundary conditions are shown in Figure \ref{sec5_fig:fracture_differentboundarycondition_sample}.

\begin{figure}[htp]
	\centering
	\includegraphics[scale=0.25]{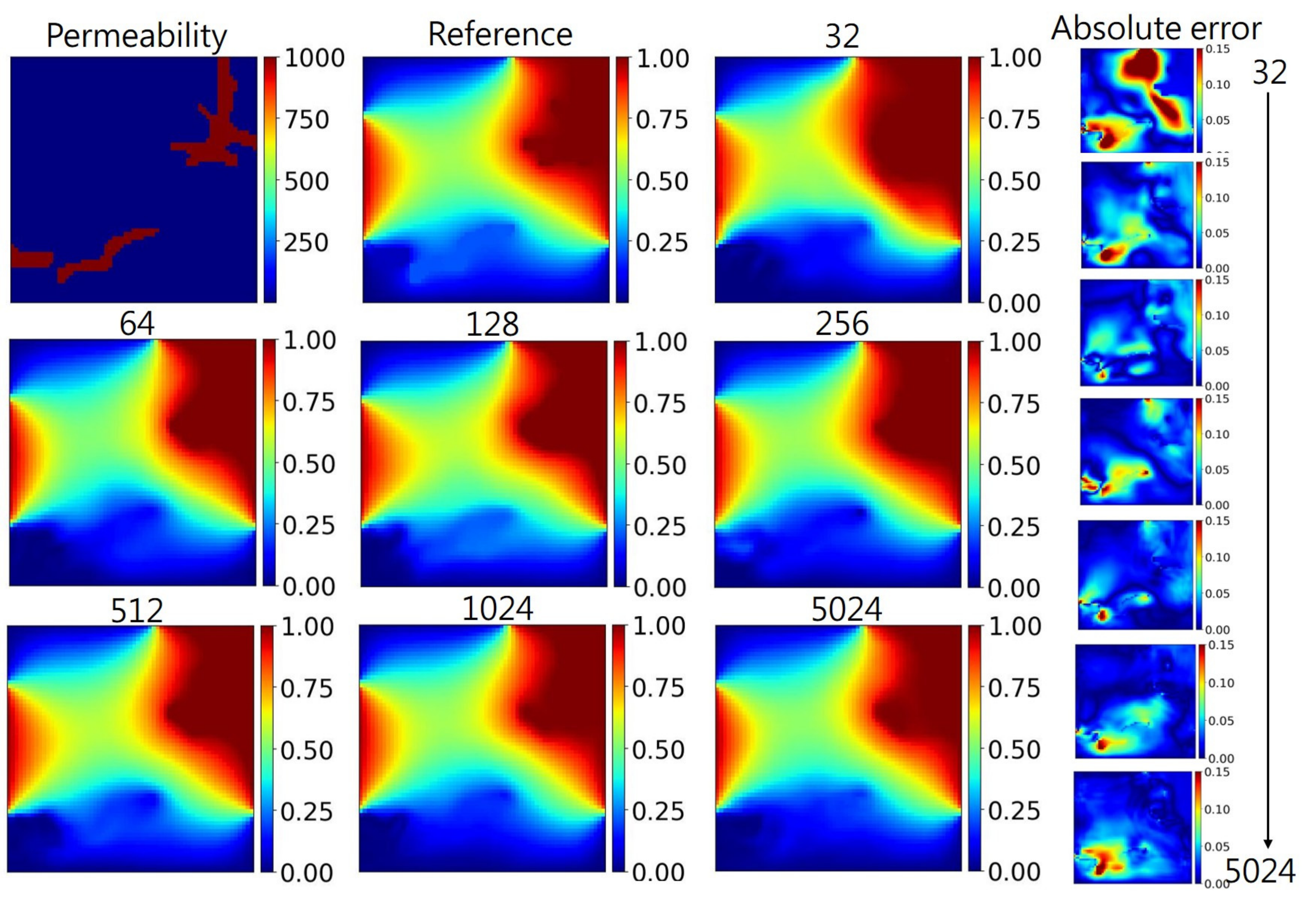}
	\caption{Fractured rocks: MITL performance on  different sizes of training dataset (TL4)}	\label{sec5_fig:fracture_bcDirichlet1_157}
\end{figure}

\begin{figure}[htp]
	\centering
	\includegraphics[scale=0.25]{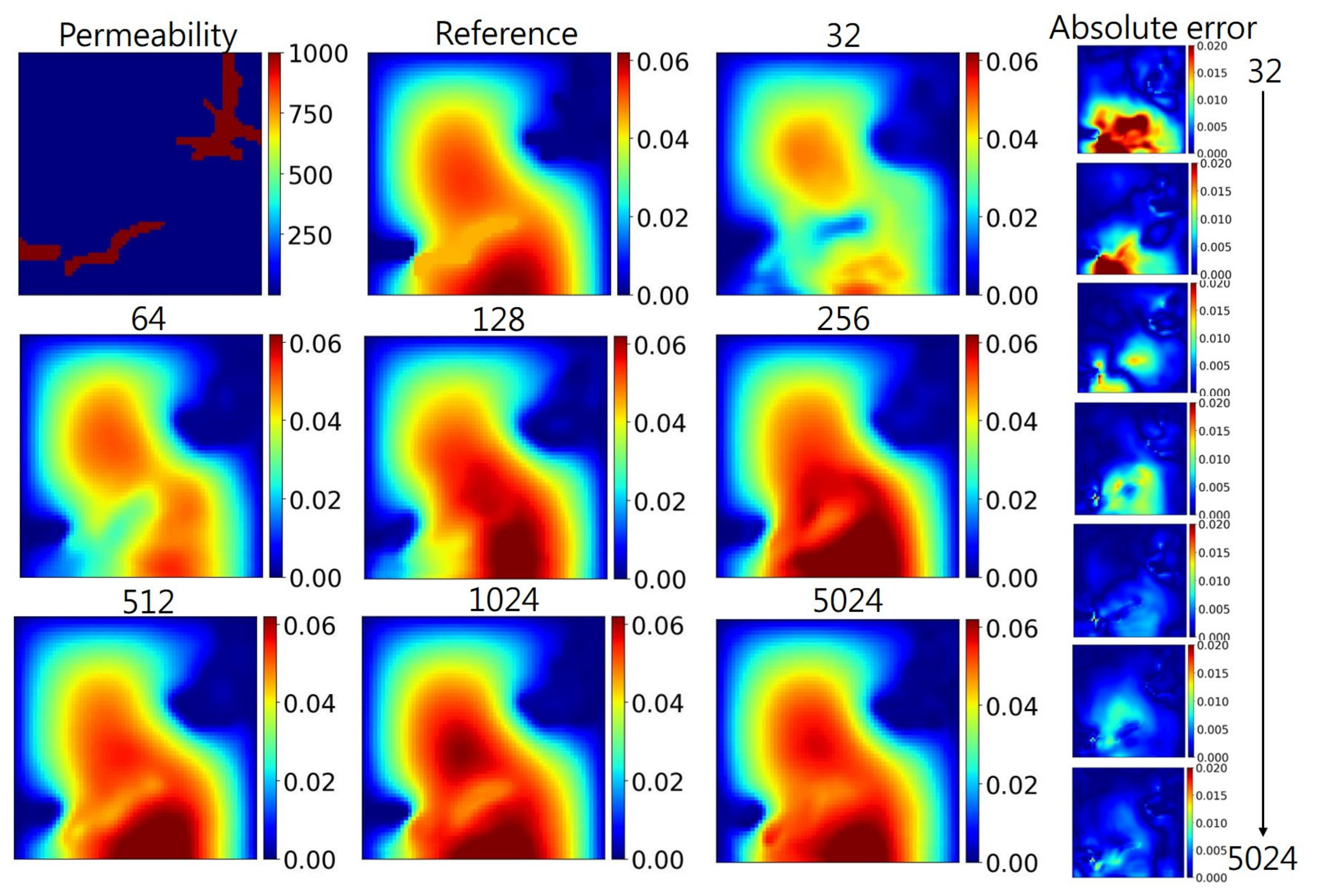}
	\caption{Fractured rocks:MITL performance on  different sizes of training dataset (TL5)}	\label{sec5_fig:fracture_bcmixed1_157}
\end{figure}

The MITL is trained for above target tasks, and the results are presented in the top two rows of Table  \ref{sec5_table:fracture_differentboundarycondition_sample}. It can be observed that the overall performance in this case is not good as that of channelized acquifers due to the complexity of random input field. However, with 128–256 training samples MITL succeeds in learning the overall solution structure. This level of accuracy meets the needs of design and optimization tasks that do not demand high fidelity (see Figure \ref{sec5_fig:fracture_bcDirichlet1_157} and Figure \ref{sec5_fig:fracture_bcmixed1_157}). Furthermore, comparing the results in the first row with those in the third and fourth rows of Table  \ref{sec5_table:fracture_differentboundarycondition_sample}, it can be seen that MITL has a significant advantage over CNN-based ROM retrained from scratch when the data size is small. In the third row, the parameter scale is consistent with that of the source model, while the fourth row shows the results after adjusting the network parameter to 264,772 (the parameter scale of MITL is 276,654).
\begin{table}[H]
	\centering  
	\setlength{\abovecaptionskip}{0cm}
	\setlength{\belowcaptionskip}{0.2cm}
	\caption{Fractured rocks:MITL performance for problems with different boundary conditions}  
	\label{sec5_table:fracture_differentboundarycondition_sample}  
	\scalebox{1.}{
		\begin{tabular}{cccccccc}
			\toprule  
			& 5024 & 1024 & 512 & 256 & 128 & 64 & 32  \\  
			\midrule
			MITL(TL4) &6.14\%& 7.92\%& 8.67\%& 9.54\%& 10.96\%& 12.55\%& 14.88\%\\
			MITL(TL5) &5.71\%& 7.03\%& 8.31\%& 10.81\%& 12.50\%& 14.30\%& 16.98\%\\		
			CNN-based ROM 1(TL4) &4.02\%& 8.05\%& 12.73\%& 14.66\%& 17.75\%& 22.50\%& 24.13\%\\	
			CNN-based ROM 2(TL4) &7.63\%& 12.06\%& 14.41\%& 16.71\%& 20.45\%& 22.52\%& 26.32\%\\				
			\bottomrule
		\end{tabular}
	}
\end{table}
\subsubsection{Transfer to problems with different operators}
In this section, we turn to a more difficult transfer task, that is, transferring the multiscale features extracted by the source model to target tasks with different differential operators. Accordding to the complexity of differential operators, we first consider reaction-diffusion process,
\begin{itemize}
	\item \textbf{[TL6] Reaction-diffusion equation}
	\begin{equation}
		\left\{
		\begin{aligned}
			&-\nabla\cdot\Big(\kappa(x,\xi) \nabla u(x)\Big)+200 u(x) = \exp(x), \quad x \in \Omega \\
			& u(x) = 0, \quad x \in \partial \Omega.
		\end{aligned}
		\right.
		\nonumber
	\end{equation}
\end{itemize}
We further consider two nonlinear operators. The first one involves a nonlinear source term, and the second is the p-Laplacian equation with $p=3$. Both operators exhibit mild nonlinearity and therefore share similarity with the source tasks. 
\begin{itemize}
	\item \textbf{[TL7] Nonlinear source term}
	\begin{equation}
		\left\{
		\begin{aligned}
			&-\nabla \cdot \Big(\kappa(x,\xi) \nabla u(x)\Big) = \sin(10\pi u(x))+cos(10\pi u(x)), \quad x \in \Omega \\
			& u(x) = 0, \quad x \in \partial \Omega.
		\end{aligned}
		\nonumber
		\right.
	\end{equation}
	\item \textbf{[TL8] P-Laplacian Equation}
	\begin{equation}
		\left\{
		\begin{aligned}
			&-\nabla \cdot \Big( \kappa(x,\xi)|u(x)|\nabla u(x)\Big) =\exp(x),\quad x \in \Omega,\\
			&u(x) =0, \quad x \in \partial{\Omega}.
		\end{aligned}
		\right.
		\nonumber
	\end{equation}
\end{itemize}

\begin{figure}[htp]
	\centering
	\includegraphics[scale=0.22]{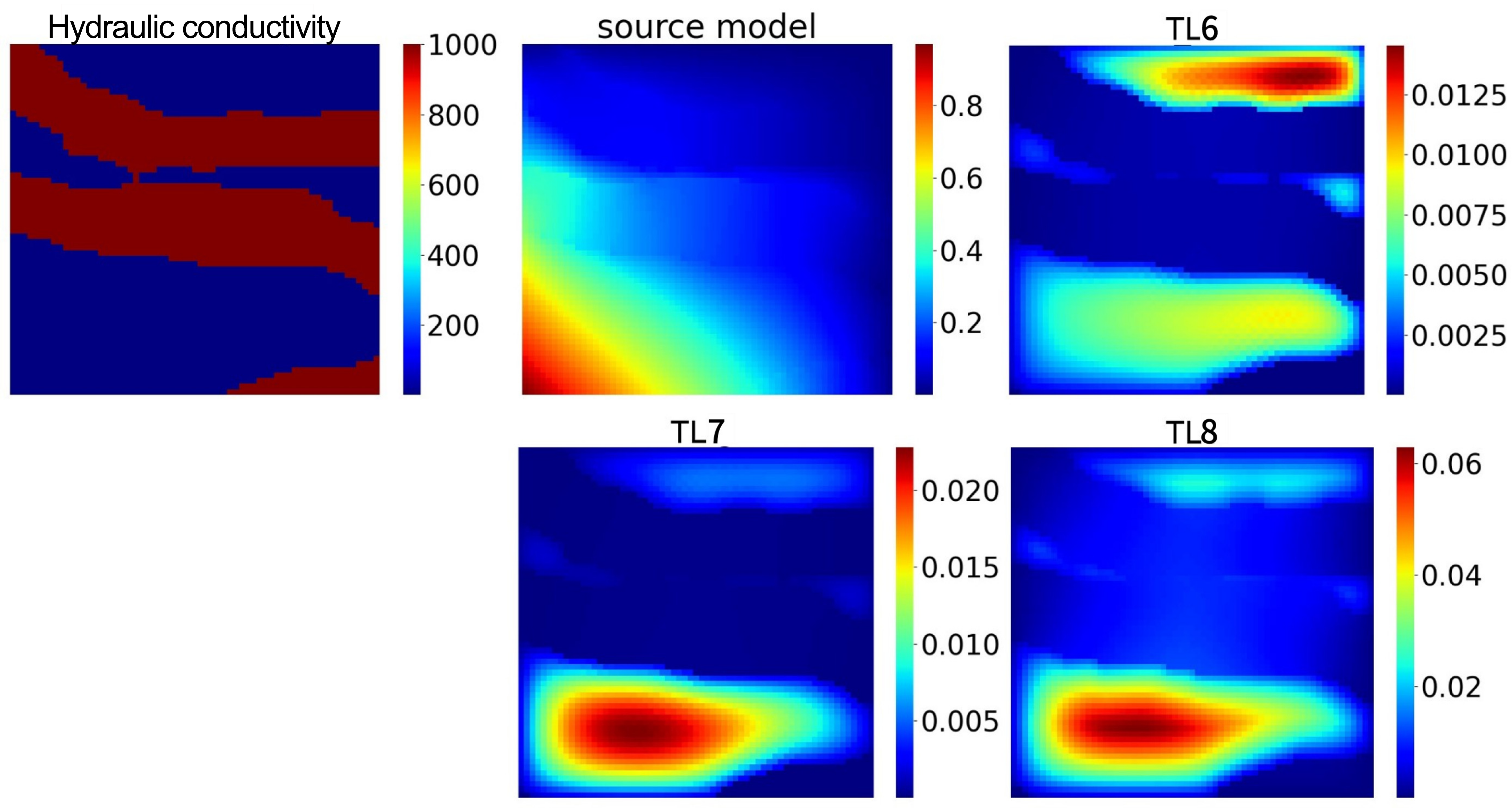}
	\caption{Channelized acquifer: solution samples of problems with three different differential operators}	\label{sec5_fig:groundwater_differentoperator_sample}
\end{figure}

In the following experiments, we consider the random input field defined in Section \ref{ssec:channelized aquifers}. For the same input sample $K$, the solutions corresponding to these three differential operators are shown in Figure \ref{sec5_fig:groundwater_differentoperator_sample}.
The target model is trained with training sets of sizes 5024, 1024, 512, 256, 128, and 64, respectively. The relative $L_2$ error is evaluated on the same test set of size 1000, and the results are presented in Table \ref{sec5_table:groundwater_differenoperator}.
\begin{table}[H]
	\centering  
	\setlength{\abovecaptionskip}{0cm}
	\setlength{\belowcaptionskip}{0.2cm}
	\caption{Channelized acquifer: MITL performance of problems with different differential operators}  
	\label{sec5_table:groundwater_differenoperator}  
	\scalebox{1.}{
		\begin{tabular}{ccccccc}
			\toprule  
			& 1024 & 512 & 256 & 128 & 64 & 32 \\  
			\midrule
			TL6 & 2.44\%& 2.93\%& 4.79\%& 5.16\%& 6.50\%& 12.27\%\\
			TL7 & 2.13\%& 2.50\%& 4.00\%& 4.56\%& 7.00\%& 12.98\%\\
			TL8 &2.31\% &2.56\%&3.47\%&4.32\%&7.58\%&10.99\%\\							
			\bottomrule
		\end{tabular}
	}
\end{table}
Above differential operators are steady, we now consider a temporal-spatial problem, namely,
\begin{itemize}
	\item \textbf{[TL9] Parabolic equation}
	\begin{equation}
		\left\{
		\begin{aligned}
			&\frac{\partial u(x,t)}{\partial t}=\nabla \cdot \Big( \kappa(x,\xi)\nabla u(x,t)\Big)-\exp(x_1+x_2),\quad x \in \Omega,\quad t\in [0,0.09]\\
			&u(x,t)=\sin(\pi x_1)+\cos(\pi x_2),\quad x \in \partial{\Omega},\\
			&u(x,0) = \sin(\pi x_1)+\cos(\pi x_2),\quad x\in \Omega.
		\end{aligned}
		\right.
		\nonumber
	\end{equation}
\end{itemize}
Using a uniform time interval $0.0015$, the temporal domain is discretized into $60$ steps for each random input sample $K$. Twelve of these time steps are allocated to training, while the rest are used to assess the method's interpolation accuracy. As illustrated in Table \ref{sec5_table:groundwater_parabolic}, MITL have a good performance in the temporal-spatial problems.
\begin{table}[H]
	\centering  
	\setlength{\abovecaptionskip}{0cm}
	\setlength{\belowcaptionskip}{0.2cm}
	\caption{Channelized acquifer:MITL performace of the parabolic problem (``Training points'' represents the prediction performance at the training points, and ``Interpolation'' denotes the prediction performance at all time steps).}  
	\label{sec5_table:groundwater_parabolic}  
	\scalebox{1.}{
		\begin{tabular}{ccccc}
			\toprule  
			& 1024 & 512 & 256 & 128 \\  
			\midrule
			TL9 (Training points) & 2.63\%& 3.01\%& 4.56\% & 5.96\%\\
			TL9 (Interpolation) & 2.75\%& 3.15\%& 4.65 & 6.01\%\\
			\bottomrule
		\end{tabular}
	}
\end{table}
\begin{figure}[htp]
	\centering
	\includegraphics[scale=0.45]{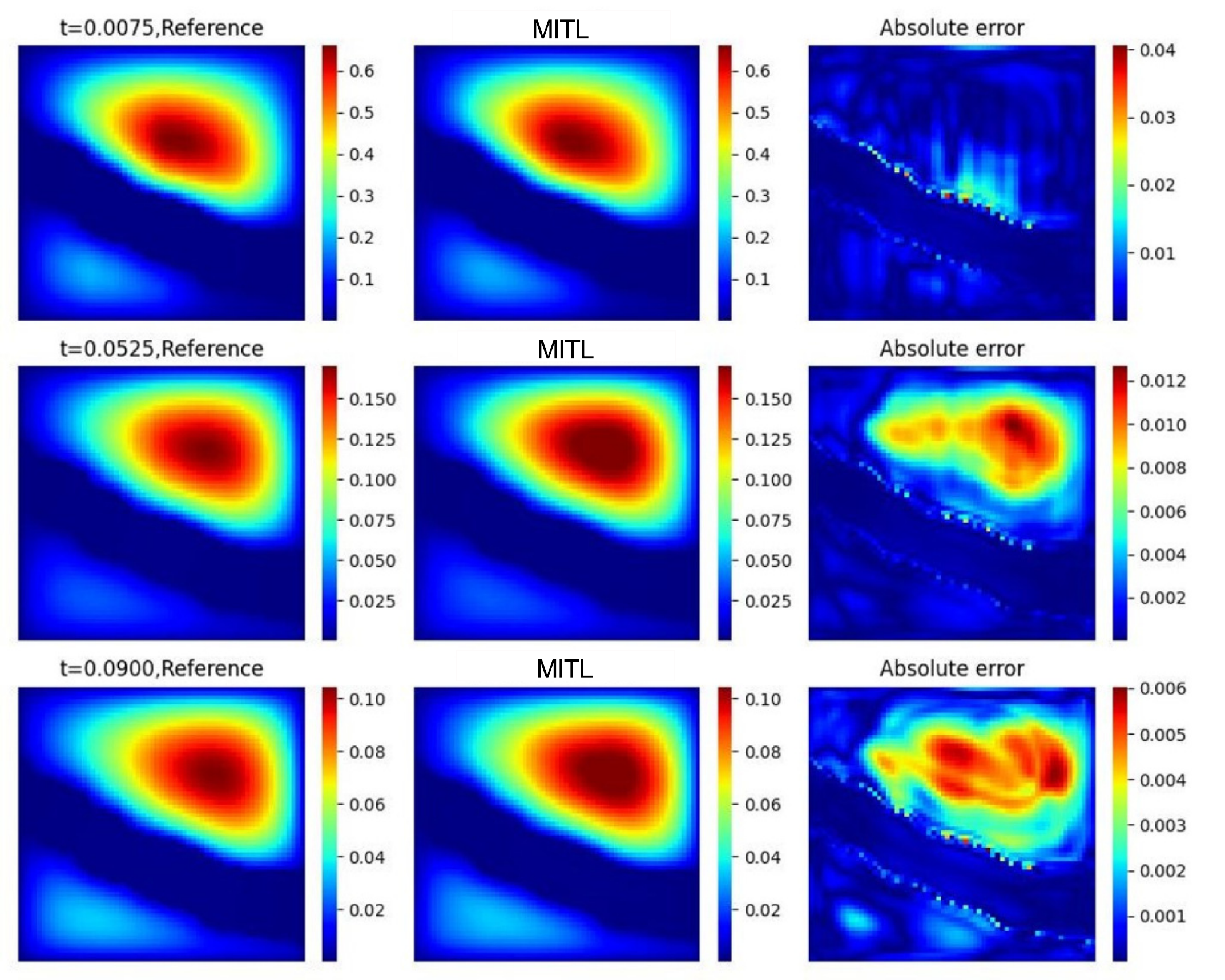}
	\caption{Channelized acquifer: MITL performance of parabolic equations (the size of training dataset is 512).}	\label{sec5_fig:groundwater_parabolic_train}
\end{figure}
To more intuitively illustrate the predictive performance across different time steps, we further plot the predicted results at some time steps, as shown in Figure \ref{sec5_fig:groundwater_parabolic_train}. It can be observed that the error decreases progressively as the time increases. This arises because the solution converges toward the steady-state limit as the time increases, making its data distribution closer to that of the source problem. 

\subsection{Application to inverse problems}
The above experimental results demonstrate that the proposed method can provide an efficient and accurate surrogate model for multiscale problems with random inputs (\ref{sec2:problem setup}) under limited data availability. In this section, we apply MITL to inverse problems. Here, we consider the multiscale systems modeled by the equations,
\begin{equation}
	\left\{
	\begin{aligned}
		&-\nabla \cdot \Big(\kappa(x,\xi)\nabla u(x)\Big) = f(x),\quad x \in \Omega = [0,1]\times[0,1],\\
		& u(x) = 1,\quad x \in \partial{\Omega},
	\end{aligned}
	\right.
	\label{sec5_eq:inverse_equations}
\end{equation}
where $\kappa(x,\xi)$ is the random input field defined in Section \ref{ssec:channelized aquifers}. And the source term $f(x)$ is defined as follows,
\begin{equation*}
	f(x) = \frac{100}{2\pi \sigma^2} \exp(\frac{\Vert x-[0.5,0.5] \Vert_2^2}{2\sigma^2}).
\end{equation*}
Take $\sigma = 0.45$ and assume that the observation noise $\sigma_{\text{obs}} = 0.05$. Then observation data can be generated by solving equation (\ref{sec5_eq:inverse_equations}) and adding Gaussian noise,
\begin{equation*}
	y(x) = u_h(x)+\epsilon,\quad \epsilon \sim \mathcal{N}(0,\sigma_{\text{obs}}).
\end{equation*}

We can thus obtain the dataset $\mathcal{Y}^{(i)} = \Big\{y(x_1),y(x_2),\cdots,y(x_{N_{\text{obs}}})\Big\},\quad i=1\cdots M$ observed by $N_{\text{obs}}$ sensors on the spatial points $\mathcal{X}_{\text{obs}} = \Big\{x_1,x_2,\cdots,x_{N_{\text{obs}}}\Big\}$, our goal is to estimate the corresponding random input sample $K_i$. The size of training dataset $M$ is less than 5024, and the number of test samples is 1000.
\begin{figure}[htp]
	\centering
	\includegraphics[scale=0.3]{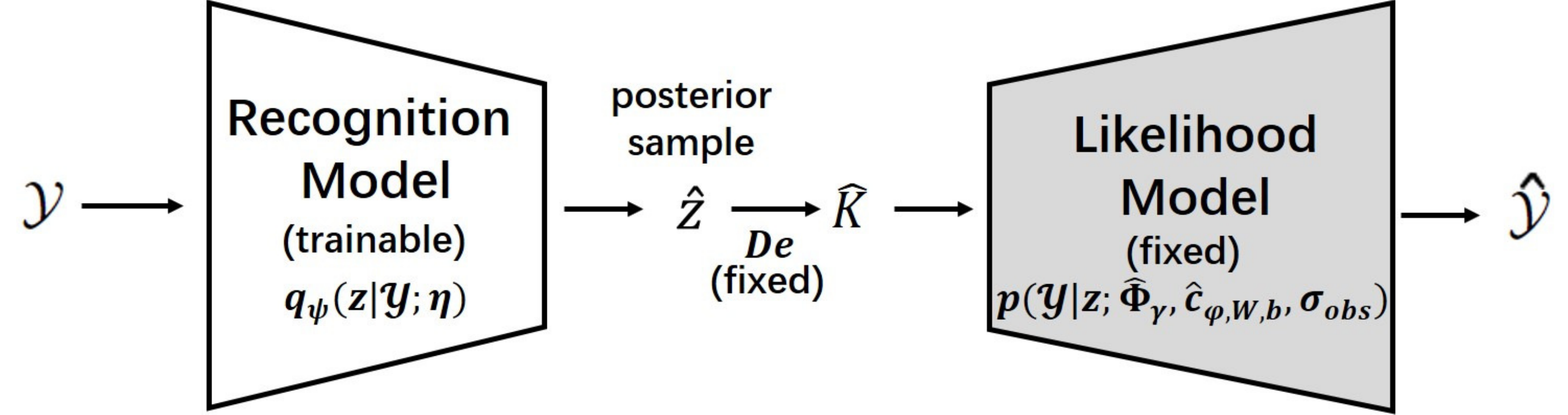}
	\caption{The schema of Surrogate-constrained VAE.}	\label{sec5_fig:VAE_outline}
\end{figure}

A method combines Variational Autoencoder (VAE) \cite{sec5:VAE} with MITL is introduced to solve this problems. VAE is a type of generative model that belongs to the family of Autoencoders, specifically designed for learning latent representation of the input data and generate new data samples. It is composed of the recognition model, which learn the mapping from input data to parameters of posteriors for latent variables, and the likelihood model, which reconstructs data from latent space. In standard VAE, both recognition model and likelihood model are represented by neural networks. As an unsupervised method, VAE learn the latent variables purely by data, which is not suitable for inverse problems. Hence, we use the trained MITL as parts of likelihood model to constrain the learning of latent variables, which is inspired by PDE-constrained Bayesian inversion methods \cite{sec5:Bayesian inverse problem}. Therefore, we call
the method Surrogate-constrained VAE (see Figure \ref{sec5_fig:VAE_outline}) and the unknown parameters $\psi$ are only contained in the recognition model, that is
\begin{equation}
	\eta = \mathcal{P}_{\psi}^{\text{Recog}}(\mathcal{Y})
\end{equation}
where $\eta$ are the local variational parameters that define the posterior distributions $q(\boldsymbol{z}|\mathcal{Y};\eta)$, which is used to approximate the true posterior $p(\boldsymbol{z}|\mathcal{Y})$. Here, the latent variable $\boldsymbol{z}$ denotes the low-dimensional representation of the random input sample $K$. We learn the corresponding $\boldsymbol{z}$ by Convolutional Autoencoder in this experiments, which can be written as,
\begin{equation}
	\hat{K} = \text{De}\circ \text{En}(K),
\end{equation}
where $\text{En}$ is the encoder that encodes $K$ to low-dimensional representation $\boldsymbol{z}$ and $\text{De}$ denotes the decoder that reconstruct $K$ from $\boldsymbol{z}$. Other efficient reduction methods, such as KLE\cite{sec5:KLE}, diffusion map\cite{sec5:diffusionmap} and so on, can also be used. The mapping $\mathcal{P}_{\psi}^{\text{Recog}}$ is defined by the standard convolutional neural networks. The problem thus becomes solving for the global variational parameters $H$. Using
such a neural network, we can obtain the posterior estimation of latent variable $z$ by passing observation data $\mathcal{Y}$ through $\mathcal{P}_{\psi}^{\text{Recog}}$, which can not be achieved by PDE-constrained Bayesian inversion methods.

Next, we will use variational inference to learn the parameters $H$. Because the likelihood of data $p(y)$ is hard to computed, we instead minimize the lower bound of it that is known as evidence lower bound (ELBO),
\begin{equation}
	\begin{aligned}
	\text{ELBO}(\psi) &= \mathbb{E}_{q_{\psi}(\boldsymbol{z}|\mathcal{Y};\eta)}\Big(\log\ p(y|\boldsymbol{z};\hat{\Phi}_{\gamma},\hat{c}_{\varphi,W,b},\sigma_{obs})\Big)\\
	&+ \text{KL}\Big(q_{\psi}(\boldsymbol{z}|\mathcal{Y};\eta)\Vert p(\boldsymbol{z})\Big),
	\label{sec5_eq:ELBO}
	\end{aligned}
\end{equation}
where $p(y|\boldsymbol{z};\hat{\Phi}_{\gamma},\hat{c}_{\varphi,W,b},\sigma_{obs})$ is the likelihood model with i.i.d Gaussian observation noise $\sigma_{obs}$, i.e., 
\begin{equation}
	\mathcal{Y}_i = \boldsymbol{M}\hat{\Phi}_{\gamma}(K_i)\hat{c}_{\varphi,W,b}(K_i)+\epsilon_i, \epsilon_i \sim \mathcal{N}(0,\sigma_{obs}^2 I),
	\nonumber
\end{equation}
where $\boldsymbol{M}=[m_{ij}]\in \mathbb{R}^{N_h\times N_h}$ is the mask matrix to control the outputs resolution, which is defined as follows,
	\begin{equation}
	m_{ij} = \left\{
	\begin{aligned}
		&1, \quad x_{ij}\in \mathcal{X}_{\text{obs}},\\
		&0, \quad \text{otherwise},
	\end{aligned}
	\right.
	\nonumber
\end{equation}
where $x_{ij}$ denotes the $i$-th row and the $j$-th column element in the spatial mesh $\mathcal{H}$. And $p(\boldsymbol{z})$ is the prior, we take it the standard Gaussian distribution. The approximation posterior distribution $q(\boldsymbol{z}|\mathcal{Y};\psi)$ are also Gaussian with mean vector $\mu$ and covariance matrix $\sigma^2 I$, i.e.,
\begin{equation}
	\boldsymbol{z}|\mathcal{Y} \sim \mathcal{N}(\mu,\sigma^2 I).
	\nonumber
\end{equation}
Furthermore, we can understand the equation (\ref{sec5_eq:ELBO}) from the perspective of inverse problems. The first term aims at maximizing the likelihood of the reconstructed data. The second term is the Kullback-Leibler divergence of two probability density, which is defined as
\begin{equation}
	\text{KL}(p \Vert q)=\int p(x) \log\frac{p(x)}{q(x)} \ dx.
	\nonumber
\end{equation} 
It works as the regularization term such that posterior is as similar as possible to prior, which can help to alleviate the ill-posedness of the inverse problem. In terms of reparametrization tricks \cite{sec5:VAE}, we can use gradient descent algorithms, such as Adam, to optimize neural network parameters $H$. Once the recognition model is trained, we can also estimate the expectations of $K$ by Monte Carlo method, i.e.,
\begin{equation}
	\mathbb{E}_{q_{\psi}(\boldsymbol{z}|\mathcal{Y};\eta)} (K) 
	\approx \frac{1}{N_{\text{MC}}}\sum_{i=1}^{N_{\text{MC}}}\text{De}(\boldsymbol{z}^{(i)}), \boldsymbol{z}^{(i)}\sim q_{\psi}(\boldsymbol{z}|\mathcal{Y};\psi).
	\label{sec5_eq:expectation}
\end{equation}
Similarly, we can also obtain the variance of $K$ based on estimations (\ref{sec5_eq:expectation})
\begin{equation}
	Var_{q_{\psi}(\boldsymbol{z}|\mathcal{Y};\eta)} (K) 
	\approx \frac{1}{N_{\text{MC}}}\sum_{i=1}^{N_{\text{MC}}} \Big( \text{De}(\boldsymbol{z}^{(i)}) -\mathbb{E}_{q_{\psi}(\boldsymbol{z}|\mathcal{Y};\eta)} (K)\Big)^2, \boldsymbol{z}^{(i)}\sim q_{\psi}(\boldsymbol{z}|\mathcal{Y};\eta).
	\label{sec5_eq:variance}
\end{equation} 

\begin{figure}[htp]
	\centering
	\includegraphics[scale=0.25]{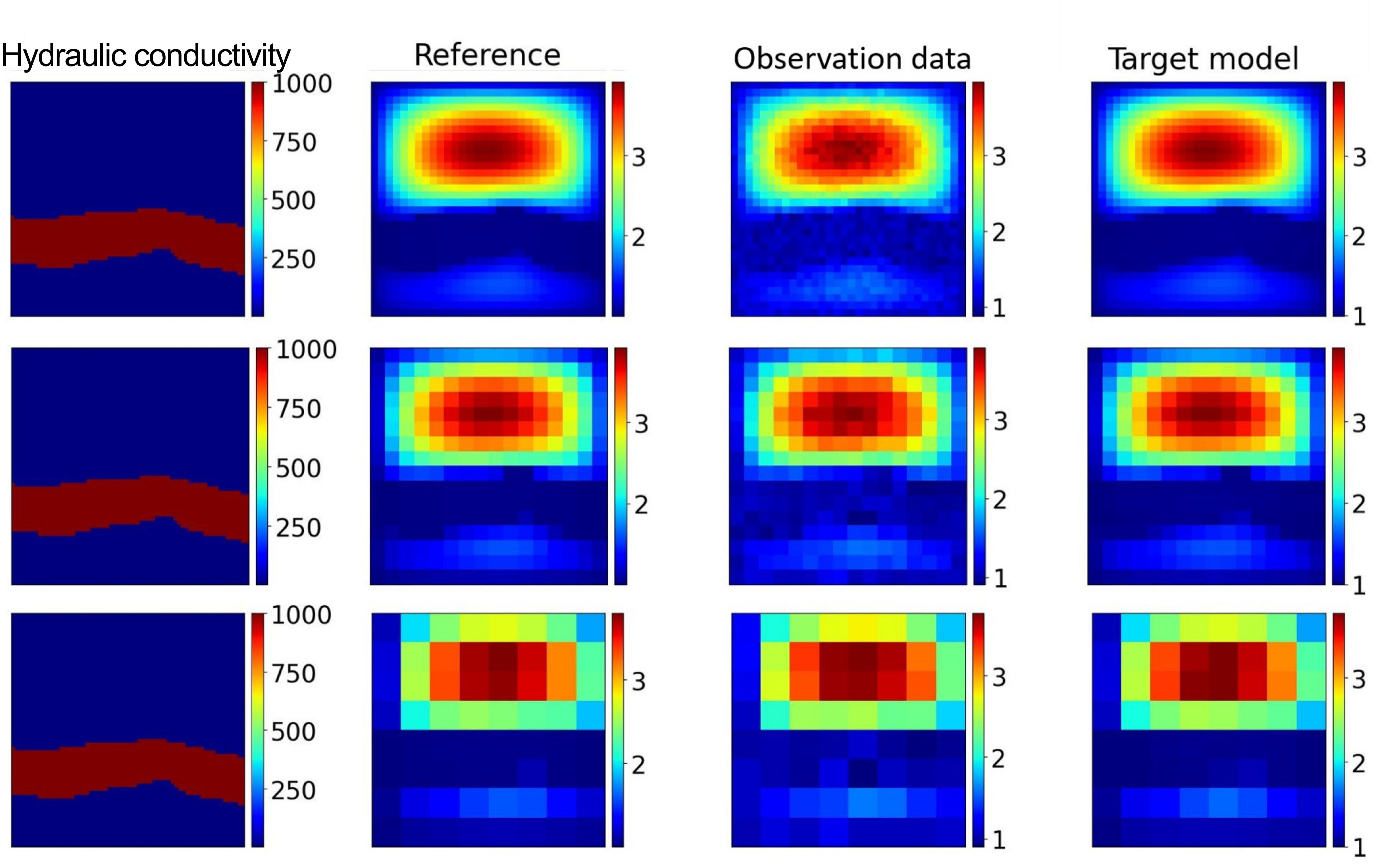}
	\caption{MITL performance on noisy data. (The resolutions are $32\times 32$, $16\times16$ and $8\times8$ from top to bottom, respectively. And the noise level $\sigma_{\text{obs}}$ is $0.05$.)}	
	\label{sec5_fig:noise_MITL}
\end{figure}

The MITL is firstly trained with a training set of size 512 on observation grids of $32\times32$, $16\times16$ and $8\times8$, respectively. The predition results are shown in Figure \ref{sec5_fig:noise_MITL}.

\begin{figure}[htp]
	\centering
	\includegraphics[scale=0.26]{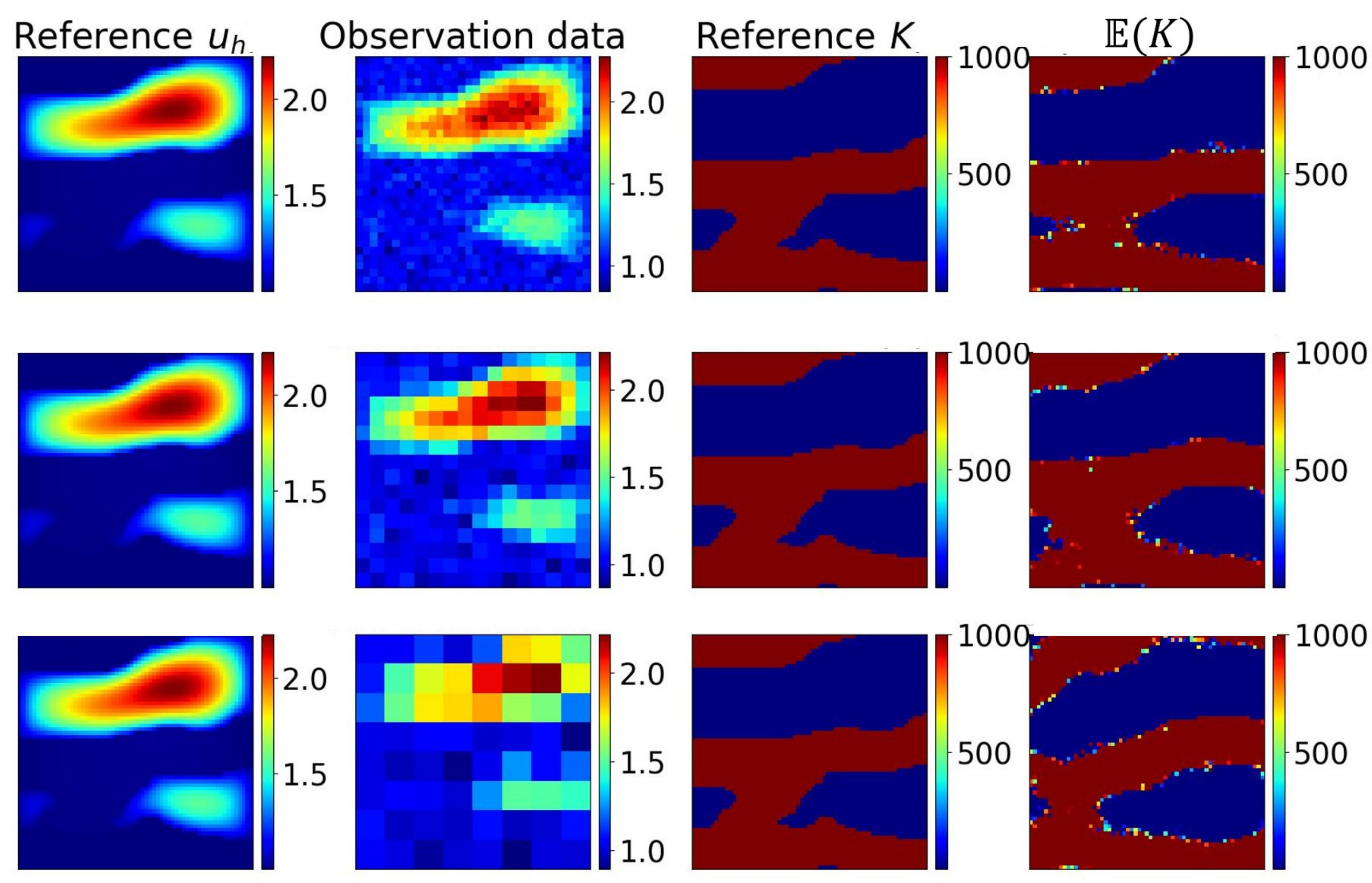}
	\caption{Inversion performance of Surrogate-constrained VAE on the test dataset.(The resolutions are $32\times 32$, $16\times16$ and $8\times8$ from top to bottom, respectively. And the noise level $\sigma_{\text{obs}}$ is $0.05$.)}	
	\label{sec5_fig:VAE_different_resolution}
\end{figure}

\begin{figure}[htp]
	\centering
	\includegraphics[scale=0.23]{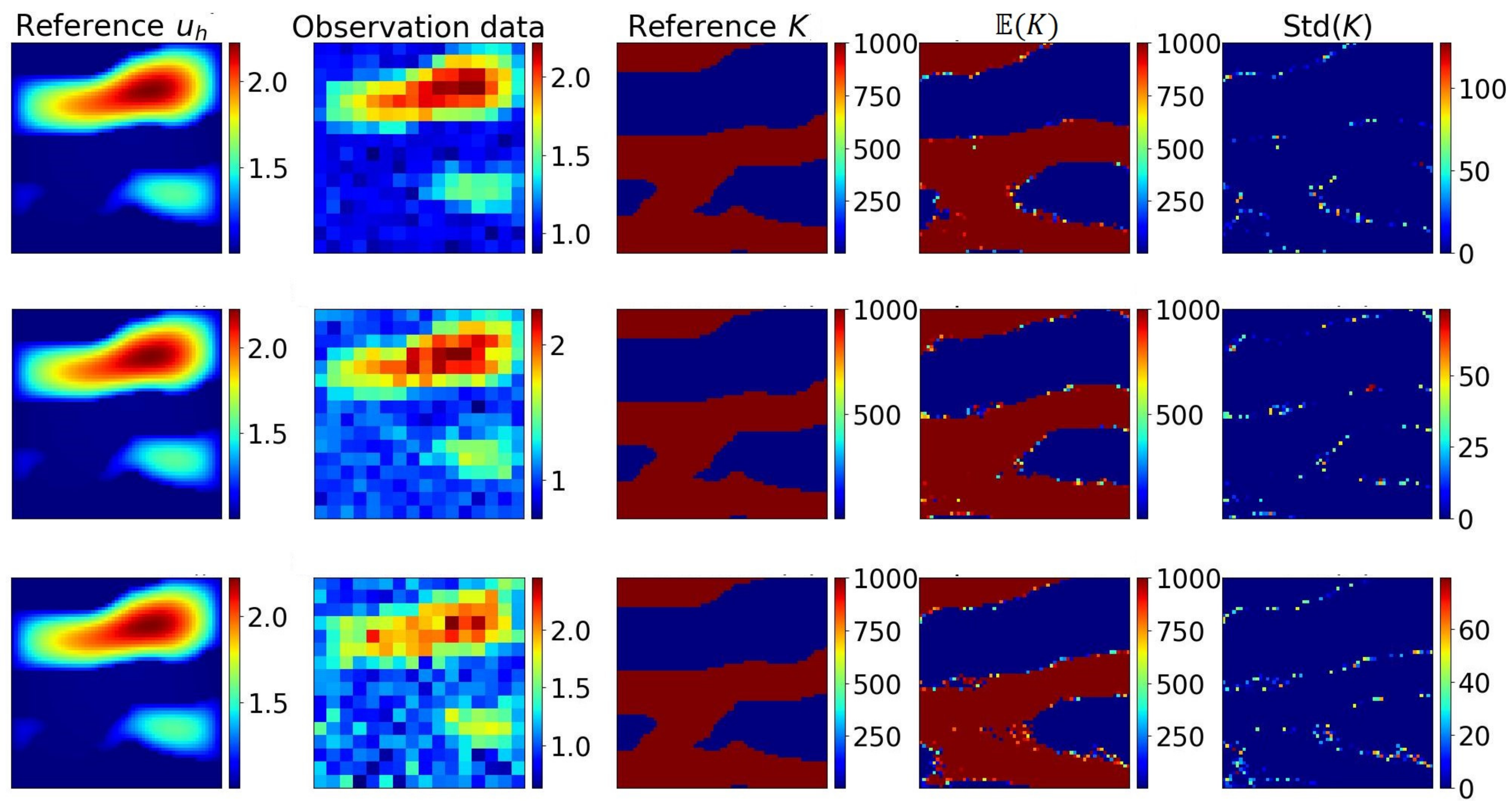}
	\caption{The uncertainty quantification ability of Surrogate-constrained VAE on the test dataset.(The noise levels are $0.05$, $0.10$ and $0.15$ from top to bottom, respectively. And the resolution is $16\times16$.)}	
	\label{sec5_fig:VAE_different_noise}
\end{figure}

The proposed Surrogate-constrained VAE is trained using the surrogates from MITL. The inversion results under different resolutions are shown in Figures \ref{sec5_fig:VAE_different_resolution}.
To further illustrate the uncertainty quantification ability of the proposed method, the resolution $16\times 16$ is unchanged and we choose the noise level as $0.05, 0.10$ and $0.15$ to train surrogate-constrained VAE. The prediction results are shown in Figure \ref{sec5_fig:VAE_different_noise}.
\section{Conclusion}
In this paper, we proposed a transfer learning strategy called MITL. It addressed three limitations of our previous CNN-based ROM: dependence on underlying PDE information, high data acquisition cost, and poor adaptability to problems with different source terms and boundary conditions. MITL is motivated by the transferability of multiscale basis functions in MsFEM. It transfers the multiscale features learned from a source task to target tasks by fine-tuning lightweight neural networks, thereby enabling efficient transfer across similar multiscale problems.

Specifically, we first reviewed the core idea of MsFEM from a transfer learning perspective. Inspired by this observation, a source task was designed in which MsFEM basis functions were solved over the global domain to preserve globally coupled multiscale information. CNN-based ROM was trained on this source task, yielding transferable representations that encodes multiscale features. For the target task, two sub-networks, TL-Basis CNNs and TL-Coef CNNs, were introduced to fine-tune the previously learned Basis CNNs and Coef CNNs, respectively. TL-Basis CNNs keep the parameters of Basis CNNs fixed and adopt a residual structure with CNOs to adjust the reduced basis learned from source tasks. The resulting mapping is a nonlinear function of the basis. Such an architecture enables MITL to handle nonlinear problems. Moreover, it makes the mapping from basis to outputs nonlinear, which is consistent with the homogenization theory of periodic multiscale problems. TL-Coef CNNs achieve rapid adaptation to both steady-state and unsteady problems by freezing the backbone, fine-tuning only the top layers, and incorporating a temporal encoding network. The trainable parameters of the entire framework account for only small portion of those of the source model, substantially improving the training efficiency on target tasks.   We note that the idea of MITL  can be also formulated using  other multiscale basis functions.

\section{Acknowledgements}
L. Jiang acknowledges the support of NSFC 12271408. EC's research is partially supported by the Hong Kong RGC General Research Fund (Project numbers: 14305624 and 14304525).


\begin{thebibliography}{99}
\bibitem{sec1:Darcy_flow_porous}M. Alotaibi, H. Chen, S. Sun, Generalized multiscale finite element methods for the reduced model of Darcy flow in fractured porous media, Journal of Computational and Applied Mathematics. 413 (2022) 114305.
\bibitem{sec1:inverse problems}Y. Xia, N. Zabaras, Bayesian multiscale deep generative model for the solution of high-dimensional inverse problems, Journal of Computational Physics. 455 (2022) 111008.
\bibitem{sec1:optimal control}L. Jiang, L. Ma, A hybrid model reduction method for stochastic parabolic optimal control problems, Computer Methods in Applied Mechanics and Engineering. 370 (2020) 113244.
\bibitem{sec1:multiscale design} J. Capers, L. Stanfield, J. Sambles, S. Bayes, A. Powell, A. Hibbins, S. Horsley, Multiscale design of large and irregular metamaterials, Physical review applied. 21 (2024) 014005.
\bibitem{sec1:POD-based ROM} R. Zimmermann, S. Görtz, Improved extrapolation of steady turbulent aerodynamics
using a non-linear POD-based reduced order model, The Aeronautical Journal. 116 (2012) 1079-1100.
\bibitem{sec1:greedy-based ROM} M. Yano, Discontinuous Galerkin reduced basis empirical quadrature procedure for
model reduction of parametrized nonlinear conservation laws, Advances in Computational Mathematics. 45 (2019) 2287-2320.
\bibitem{sec1:MsFEM1} E. Chung, Y. Efendiev, T. Hou, Multiscale model reduction: multiscale finite element methods and their generalizations, Springer, Switzerland, 2023. 
\bibitem{sec1:MsFEM2}T. Hou, X. Wu, A multiscale finite element method for elliptic problems in composite materials and porous media. Journal of Computational Physics. 134 (1997) 169-189.
\bibitem{sec1:MsFEM3} L. Jiang, Y. Efendiev, G. Victor, Multiscale methods for parabolic equations with continuum spatial scales, Discrete and Continuous Dynamical Systems. 8 (2012) 833-859.
\bibitem{sec1:GMsFEM1} Y. Efendiev, J. Galvis, T. Hou,  Generalized multiscale finite element methods (GMsFEM), Journal of Computational Physics. 251 (2013) 116-135. 
\bibitem{sec1:CEM-GMsFEM1} M. Li, E. Chung, L., Jiang, A constraint energy minimizing generalized multiscale finite element method for parabolic equations, Multiscale Modeling and Simulation. 17 (2019) 996-1018.
\bibitem{sec1:Splitting}L. Jiang, P. Michael, A resourceful splitting technique with applications to deterministic and stochastic multiscale finite element methods, Multiscale modeling and simulation 10 (2012) 954-985.
\bibitem{sec1:intrusive_ROM1}M. Raissi, P. Perdikaris, G.E. Karniadakis,
Physics-informed neural networks: a deep learning framework for solving forward and inverse problems involving nonlinear partial differential equations.
Journal of Computational Physics, 378 (2019) 686-707.
\bibitem{sec1:intrusive_ROM2}W. Chen, Q. Wang, J. Hesthaven, C. Zhang,
Physics-informed machine learning for reduced-order modeling of nonlinear problems.
Journal of Computational Physics. 446 (2021), 110666.
\bibitem{sec3_CNNbasedROM}X. Zhang, L. Jiang, CNN-based reduced-order modeling for multiscale problems, Journal of Computational Physics. 2025, (524) 113710.
\bibitem{sec1:DeepGlobalModelReduction} S. Cheung, E. Chung, Y. Efendiev, E. Gildin, Y. Wang, J. Zhang, Deep global model reduction learning in porous media flow simulation, Computational Geosciences. 24 (2020) 261-274.
\bibitem{sec1:ROMs1} J. Hesthaven, S. Ubbiali, Non-intrusive reduced order modeling of nonlinear problems using neural networks, Journal of Computational Physics. 363 (2018) 55-78.
\bibitem{sec1:ROMs2} S. Fresca, L. Dede, A. Manzoni, A comprehensive deep learning-based approach to reduced order modeling of nonlinear time-dependent parametrized PDEs, Journal of Scientific Computing. 87 (2021) 1-36.
\bibitem{sec1:ROMs3}M. Mignolet, A. Przekop, S. Rizzi, S. Spottswood, A review of indirect/non-intrusive reduced order modeling of nonlinear geometric structures, Journal of Sound and Vibration. 332 (2013) 2437-2460.
\bibitem{sec1:ROMs4}Q. Wang, J. Hesthaven, D. Ray, Non-intrusive reduced order modeling of unsteady flows using artificial neural networks with application to a combustion problem, Journal of Computational Physics. 384 (2019) 289-307.
\bibitem{sec1:ROMs5}X. Zhang, L. Jiang, Conditional variational autoencoder with Gaussian process regression recognition for parametric models, Journal of Computational and Applied Mathematics. 438 (2024) 115532.
\bibitem{sec1:TL1} T. Wang, J. Bai, Z. Lin, Q. Wang, C. Anitescu, J. Sun, M. Eshaghi, Y. Gu, X. Feng, X. Zhuang, T. Rabczuk, Artificial intelligence for partial differential equations in computational mechanics: A review, Applied Mechanics Reviews. (2024) 1-81.
\bibitem{sec1:TL2} L. Yang, S. Liu, T. Meng, S. J. Osher, In-context operator learning with data prompts for differential equation problems, Proceedings of the National Academy of Sciences. 120 (2023) e2310142120.
\bibitem{sec1:TL3} Y. Gao, K. C. Cheung, M. K. Ng, Svd-pinns: Transfer learning of physics-informed neural networks via singular value decomposition, in: 2022 IEEE Symposium Series on Computational Intelligence (SSCI). (2022) 1443–1450.
\bibitem{sec1:TL4} Z. Li, N. Kovachki, K. Azizzadenesheli, B. Liu, K. Bhattacharya, A. Stuart, A. Anandkumar, Fourier neural operator for parametric partial differential equations. ICLR, (2021).
\bibitem{sec4_nonlocal}Q. Du, B. Engquist, X. Tian, Multiscale modeling, homogenization and nonlocal effects: Mathematical and computational issues, Contemporary Mathematics. 754 (2020) 115-139.
\bibitem{sec4_CNO}B. Raonic, R. Molinaro, T. Rohner, S. Mishra and E. Bezenac, Convolutional neural operators,
ICLR 2023 workshop on physics for machine learning. 2023.
\bibitem{sec4_CNN}M. Zeiler, R. Fergus, Visualizing and understanding convolutional networks, European conference on computer vision. (2024) 818-833.
\bibitem{sec4:homogenization} G. Allaire, B. Robert, A multiscale finite element method for numerical homogenization, Multiscale Modeling sand Simulation. 4 (2005)  790-812.
\bibitem{sec5_K_groundwater}E. Laloy, R. Herault, D. Jacques and N. Linde, Training-image based geostatistical inversion using a spatial generative adversarial neural network, Water Resources Research. 54 (2018) 381-406.
\bibitem{sec5_K_fracture}M Yaqoob, M Ishaq, MY Ansari, VRS Konagandla, TA Tamimi, S Tavani, A. Corradetti, T. Seers, GeoCrack: a high-resolution dataset of fracture edges in geological outcrops, Scientific Data. 11 (2024) 1318.
\bibitem{sec5:VAE}L. Cinelli, M. Marins, E. Silva, S. Netto, Variational methods for machine learning with applications to deep networks, Springer, 2021, pp. 151-152.
\bibitem{sec5:Bayesian inverse problem}J. Cockayne, C. Oates, T. Sullivan, M. Girolami, Probabilistic numerical methods for PDE-constrained Bayesian inverse problems, AIP Conference Proceedings. 1853 (2017) 060001.
\bibitem{sec5:KLE}Y. Zhu, N. Zabaras, Bayesian deep convolutional encoder–decoder networks for surrogate modeling and uncertainty quantification, Journal of Computational Physics. 366 (2018) 415-447.
\bibitem{sec5:diffusionmap}I. Kalogeris, V. Papadopoulos, Diffusion maps-aided neural networks for the solution of parametrized PDEs, Computer Methods in Applied Mechanics and Engineering. 376 (2021) 113568.
\end{thebibliography}
\end{document}